\documentclass[12pt]{article}
\usepackage{amsmath}
\usepackage{graphicx}
\usepackage{enumerate}
\usepackage{natbib}
\usepackage{url} 

\usepackage{amssymb}
\usepackage{amsthm}
\usepackage[flushleft]{threeparttable}
\usepackage{caption}
\usepackage{mathrsfs}
\usepackage{amscd}
\usepackage[pdftex]{hyperref}
\usepackage{multirow}
\usepackage{comment}
\usepackage{array}
\usepackage{indentfirst}
\usepackage{enumitem}
\usepackage{booktabs}
\usepackage{multirow}
\usepackage{titlesec}
\usepackage{xcolor}
\usepackage{subfigure} 
\usepackage{bm}
\usepackage[titletoc]{appendix}

\newcommand{\blind}{0}

\newtheorem{lemma}{\indent L{\scriptsize{EMMA}}}
\newtheorem{remark}{\indent R{\scriptsize{EMARK}}}
\newtheorem{theorem}{\indent T{\scriptsize{heorem}}}
\newtheorem{Theorem}{\noindent T{\scriptsize{heorem}}}
\newcommand{\PreserveBackslash}[1]{\let\temp=\\#1\let\\=\temp}

\theoremstyle{definition}

\newtheorem{cor}{\indent Corollary}

\newcommand{\bas}{\begin{eqnarray*}}
	\newcommand{\eas}{\end{eqnarray*}}
\newcommand{\ba}{\begin{eqnarray}}
\newcommand{\ea}{\end{eqnarray}}
\newcommand{\bit}{\begin{itemize}}
	\newcommand{\eit}{\end{itemize}}

\newcommand{\ols}[1]{\mskip.5\thinmuskip\overline{\mskip-.5\thinmuskip {#1} \mskip-.5\thinmuskip}\mskip.5\thinmuskip} 

\newcommand{\pkg}[1]{{\fontseries{b}\selectfont #1}} 
\newcommand\PRC{{\bm{H}}}

\hypersetup{
	colorlinks=true,
	linkcolor=black,
	citecolor=black,
	urlcolor=blue,
}

\begin{document}

\def\spacingset#1{\renewcommand{\baselinestretch}%
{#1}\small\normalsize} \spacingset{1}


\if0\blind
{
  \title{\vspace{-1.5cm}\bf Spatially Clustered Varying Coefficient Model}
  \author{{\small Fangzheng Lin$^1$, Yanlin Tang$^2$, Huichen Zhu$^3$ \footnote{Corresponding author, email: hczhu@ust.hk.}\ \ and  Zhongyi Zhu$^1$}\\
  	{\small {\it $^1$Department of Statistics, Fudan University, China}}\\
  	{\small {\it $^2$Key Laboratory of Advanced Theory and Application in Statistics and Data Science - MOE,}}\\ 
  	{\small {\it School of Statistics, East China Normal University, China}}\\
  	{\small {\it $^3$Department of Mathematics, Hong Kong University of Science and Technology, China}}}
  \date{}
  \maketitle
  \vspace{-0.7cm}
} \fi

\if1\blind
{
  \bigskip
  \bigskip
  \bigskip
  \begin{center}
    {\LARGE\bf Spatially Clustered Varying Coefficient Model}
\end{center}
  \medskip
} \fi

\bigskip
\begin{abstract}
\spacingset{1.5}
In various applications with large spatial regions, the relationship between the response variable and the covariates is expected to exhibit complex spatial patterns.
We propose a spatially clustered varying coefficient model, where the regression coefficients are allowed to vary smoothly within each cluster but change abruptly across the boundaries of adjacent clusters, {and we develop a unified approach for simultaneous coefficient estimation and cluster identification. The varying coefficients are approximated by penalized splines, and the clusters are identified through a fused concave penalty on differences in neighboring locations, where the spatial neighbors are specified by the minimum spanning tree (MST). The optimization is solved efficiently based on the alternating direction method of multipliers, utilizing the sparsity structure from MST.} {
Furthermore, we establish the oracle property of the proposed method considering the structure of MST.} 
Numerical studies show that the proposed method can efficiently incorporate spatial neighborhood information and automatically detect possible spatially clustered patterns in the regression coefficients. 
An empirical study in oceanography illustrates that the proposed method is promising to provide informative results.
\end{abstract}

\noindent%

\spacingset{1.5} 
\section{Introduction}
\label{sec:intro}

With the development of remote sensors, satellites and geographic software, spatial data from large region are increasingly collected in recent years. For instance, in the motivating water mass analysis in Section \ref{real-analysis}, the data are collected over Southern Hemisphere's oceans,  and we aim to investigate the complex relationship between temperature and salinity (T-S relationship) over this large region, {which plays an important role in the ocean current and global climate system (\citeauthor{emery2001water}, \citeyear{emery2001water}; \citeauthor{nandi2004seismic}, \citeyear{nandi2004seismic})}. 
To model the T-S relationship over this large region, we have at least two main challenges. 
First, the Southern Hemisphere's oceans consist of several water masses, and 
due to the nonlinear nature of geophysical fluid dynamics \citep{vallisatmospheric}, the T-S relationship is likely to change rapidly across the narrow boundaries (termed as $fronts$ in geoscience) between adjacent water masses \citep{li2019spatial}. This phenomenon is ubiquitous in the ocean and the atmosphere, and it automatically leads to a spatially clustered pattern in the T-S relationship. 
Second, each valid water mass generally occupies a big region \citep{emery2001water}, though not as large as the whole Southern Hemisphere's oceans. {The regression coefficients within a big region usually vary across different locations \citep{propastin2008application, noresah2009modelling},
 thus the T-S relationship is expected to vary within each water mass. 
}

Existing literature only partially deals with the first or the second challenge mentioned above. To model the relationship between the response variable and covariates {over a region of interest}, spatial regression models \citep{cressie1992statistics} and spatial generalized linear regression models \citep{diggle1998model} are widely used, where the coefficients of explanatory variables are usually assumed to be constant over the whole region. However, {such constant assumption is known to be restrictive over a large region, where the regression coefficients are expected to vary \citep{finley2011comparing}, and/or possibly form spatially clustered pattern \citep{li2019spatial}. Among the existing literature, many methods were developed to capture the spatially-varying pattern of the regression coefficients, i.e., address the second challenge, while literature addressing the first challenge is relatively sparse.}
 To capture the spatially-varying pattern in the second challenge,   the geographically weighted regression (GWR) \citep{fotheringham2003geographically} and spatially-varying coefficient models (SVC) \citep{gelfand2003spatial} are two popular methods. The GWR fits a local weighted regression model at each observation, where the weight matrix is defined by a kernel function. In the SVC method,  spatially-varying coefficients are modeled as a multivariate spatial Gaussian process and then fitted into the Bayesian framework with some prior distributions. Other methods to capture the spatially-varying pattern can be found in  \cite{opsomer2008non}, \cite{lu2009adaptively}, \cite{sangalli2013spatial},  \cite{mu2018estimation}.  The main drawback of these methods is that they can not deal with the possible spatially clustered pattern, i.e., the first challenge, which may appear in practice \citep{talley2011descriptive}.
{Limited work} has been done on capturing the spatially clustered pattern in the first challenge. Recently, \cite{li2019spatial} developed a spatially clustered coefficient (SCC) regression, which uses fused LASSO \citep[least absolute shrinkage and selection operator]{tibshirani1996regression} to automatically detect spatially clustered patterns in the regression coefficients. However, the SCC method requires that the values of regression coefficients to be constant within each cluster, thus fails to address the second challenge. {
Such constant restriction can lead to massive identified clusters when the true regression coefficients vary within each cluster, see the simulation studies	in Section \ref{Smooth-varying}. In  the motivating water mass analysis in Section \ref{real-analysis}, the SCC method also identifies massive clusters in the T-S relationship, strongly suggesting that the T-S relationship may vary within each cluster,   see Figure \ref{figure6}(ii) in Section \ref{real-analysis} for more details.}

In this paper, we propose a spatially clustered varying coefficient model (SCVCM) to address both challenges discussed above,  which can not only model the spatially clustered pattern, but also allow  spatially-varying relationship between the response and the covariates within each subregion.
{To address the SCVCM}, we adopt the penalized splines (P-splines) to model the spatially-varying coefficients, and apply fused penalties to encourage homogeneity  between spline coefficient vectors at any two locations connected in an edge set, so that  the spatially clustered pattern can be detected. {The selection of the edge set should incorporate spatial neighborhood information of regression coefficients, i.e., coefficients at proximate locations are likely to be similar, possibly resulting from similar conditions for small area \citep{finley2011comparing}. 	
Inspired by \cite{li2019spatial}, we use  minimum spanning tree (MST)  to construct the edge set, where two locations are connected by the edge only when they are close in space, so that the spatial neighborhood information is utilized.} {
 Moreover,   the number of corresponding penalized terms based on MST  is small. Utilizing such property,   we develop an  efficient algorithm to solve the optimization problem, based on the alternating direction method of multipliers (ADMM).} { Furthermore, in our theoretical investigation, we establish the oracle property of the proposed method considering the structure of MST;} { namely, for any two locations connected by MST,} the proposed method works as well as we know whether they belong the same cluster or not. {To our best knowledge, such theoretical results are novel, providing important insights about the influence of MST on cluster recovering,  see details in Section \ref{asymptotics}, and they may provide theoretical support when applying MST to other various models in the future. }

Compared to the most relevant SCC method in \cite{li2019spatial}, the proposed approach has the following major differences and advantages. 
{First, the proposed approach relaxes the constant restriction on the relationship between the response and covariates within each subregion, allowing spatial variability within each subregion, which is more reasonable in investigating the T-S relationship as discussed above the third paragraph, as well as other applications, see \cite{wheeler2009comparing} and references therein.
}
Moreover, it is worth to point out that, within each subregion, the SCVCM  degenerates to a commonly-used spatially-varying model  \citep{fotheringham2003geographically},
but the SCC model degenerates to a simple linear regression model, which is often unreasonable in spatial analysis \citep{lloyd2010local}.
{Second, the SCC method is based on the fused LASSO penalty, which may not be able to correctly recover the clusters \citep{leng2006note}. 
In the proposed method, the penalties are taken to be some commonly-used concave penalties, say SCAD \citep[smoothly clipped absolute deviation]{fan2001variable} and MCP \citep[minimax concave penalty]{zhang2010nearly}, which are known to result in better performance than LASSO in cluster recovering. {Lastly,      we establish the oracle property  considering the structure of MST, which 
 provides important insights about the influence of MST on cluster recovering.}}

The proposed approach can be regarded as a model-based clustering method, which aims at detecting the spatially clustered pattern.  Among the literature, \cite{ma2017concave}, \cite{zhang2019quantile} and \cite{ZhangYingying} proposed to identify subgroups for subjects, e.g.,  patients. {All these methods are based on pairwise fused penalties, which are not suitable for the  spatial data as pairwise construction totally ignores the spatial neighborhood information, resulting in massive redundant penalties.} \cite{tibshirani2005sparsity} estimated homogeneous effects of covariates, based on fused penalties on successive differences of regression coefficients, which is not applicable to spatial data, as they do not have a natural order. \cite{ke2015homogeneity} also pursued the homogeneous effects of covariates by adopting fused penalties based on a coefficient order from preliminary estimates, which are estimated from  independent replicates, usually unavailable for spatial data.

The rest of the paper is organized as follows. We present SCVCM,  its estimating method,  and a computationally efficient algorithm in Section \ref{model-method}, the asymptotic properties in Section \ref{asymptotics}. We assess the finite sample performance of the proposed method by extensive simulation studies in Section \ref{simulation-study}, and apply the proposed method to the  water mass analysis in Section \ref{real-analysis}. Technical details are provided in the online Supplement.

\vspace{-0.6cm}
\section{Model and method}
\label{model-method}

\vspace{-0.5cm}
\subsection{Background}
\vspace{-0.3cm}
Suppose the spatial data $\{(\bm{X}(\bm{s}_i), y(\bm{s}_i)), i=1, \cdots, n\}$ are observed at locations $\bm{s}_1, \cdots, \bm{s}_n$ $\in \mathcal{D}$, where $\mathcal{D}\subset \mathbb{R}^2$ is the region of interest,   and the covariates $\bm{X}(\bm{s}_i)=\left(x_1{(\bm{s}_i)}, \cdots, x_p{(\bm{s}_i)}\right)^T$ with $x_1(\bm{s}_i)=1$. { These locations are assumed to be fixed, which is a feature of common spatial data, such as  geostatistical data and lattice data \citep{schabenberger2017statistical}.}
A commonly-used  spatially-varying regression model 	\citep{fotheringham2003geographically, opsomer2008non} is
\begin{equation}
y(\bm{s}_i)=\sum_{k=1}^p x_k(\bm{s}_i)\beta_{k}(\bm{s}_i)+\epsilon(\bm{s}_i),
\label{eqn1}
\end{equation}
where the regression coefficient $\beta_{k}(\bm{s}_i)$ is the value of a smooth function $\beta_k(\bm{s})$ at location $\bm{s}_i$,  and $\{\epsilon(\bm{s}_i)\}_{i=1}^n$ are independent random errors with mean 0 and variance $\sigma^2$; the spatial dependence in model (\ref{eqn1}) is usually assumed to be captured through the spatially-varying intercept \citep{finley2011comparing}. However, the model (\ref{eqn1}) does not consider the possible spatially clustered pattern, which exists in many applications \citep{talley2011descriptive}.

\vspace{-0.4cm}
\subsection{SCVCM and its estimation}
\vspace{-0.3cm}
{In this paper,  we propose SCVCM to model the spatially clustered pattern. Let $\{\mathcal{D}_{k}^1, \cdots, \mathcal{D}_k^{G_k}\}$ represent the $G_k$ disjoint subregions for the $k$-th covariate, satisfying $\mathcal{D}=\mathcal{D}_{k}^1\cup\cdots\cup\mathcal{D}_{k}^{G_k}$,  $k=1, \cdots, p$. 
Then,  SCVCM is defined as
\begin{equation}
y(\bm{s}_i)=\sum_{k=1}^p \sum_{g_k=1}^{G_k} x_k(\bm{s}_i)  \beta_{k}^{g_k}(\bm{s}_i)I(\bm{s}_i\in \mathcal{D}_k^{g_k})+\epsilon(\bm{s}_i), 
\label{eqnnn5}
\end{equation}
where $I(\cdot)$ is the indicator function, and  $\beta_{k}^{g_k}(\bm{s}_i)$ is the value at location $\bm{s}_i$ of an unknown smooth  function  $\beta_{k}^{g_k}(\bm{s})$ over $\mathcal{D}_k^{g_k}$.}
The assumption of $\epsilon(\bm{s}_i)$ is the same as  in model (\ref{eqn1}).
Model (\ref{eqnnn5}) is a generalization of the model (\ref{eqn1}), allowing spatially clustered patterns for regression coefficients, which  has two important features. (i) Similar to model (\ref{eqn1}),  it allows the associations between the response and covariates to exhibit smooth variation within each subregion. (ii) It allows the investigation of different clustered patterns in different regression coefficients.

We start from an ideal case, where $\{\mathcal{D}_k^{g_k}, g_k=1, \cdots, G_k\}$, $k=1, \cdots, p$, are known.  To estimate $\beta_{k}^{g_k}(\bm{s})$ , $\bm{s}\in  \mathcal{D}_k^{g_k}$,   we  adopt the $P$-spline method, which is popular for modeling smooth variations in the context of spatial statistics \citep{ruppert2003semiparametric}.  To be specific,   in the context of $P$-splines,    it assumes that $\beta_{k}^{g_k}(\bm{s})$ , $\bm{s}\in  \mathcal{D}_k^{g_k}$,  can be approximated sufficiently well by
$(\bm{a}^{g_k}_k)^T\bm{B}(\bm{s})$,
where $\bm{a}^{g_k}_k\in \mathbb{R}^L$ is the spline coefficient vector, and $\bm{B}(\bm{s})=\left(B_1(\bm{s}), \cdots, B_L(\bm{s})\right)^T$ is the  basis function vector constructed by a large number of knot locations, see details in Section \ref{basis-function}.  As is commonly done in the $P$-spline context \citep{ruppert2003semiparametric, opsomer2008non}, we assume that $L$ is large and fixed, and the lack-of-fit error $\beta_{k}^{g_k}(\bm{s})-(\bm{a}^{g_k}_k)^T\bm{B}(\bm{s})$ is negligible uniformly over $\bm{s}\in \mathcal{D}_k^{g_k}$, so that we can simply take $\beta_{k}^{g_k}(\bm{s})=(\bm{a}^{g_k}_k)^T\bm{B}(\bm{s})$.    Then,   estimating $\beta_{k}^{g_k}(\bm{s})$ for $\bm{s}\in \mathcal{D}_k^{g_k}$ is equivalent to estimate $\bm{a}^{g_k}_k$.
Moreover, we prove that,  $\bm{d}_1^T\bm{B}(\bm{s})\neq \bm{d}_2^T\bm{B}(\bm{s})$ for some $\bm{s}\in\mathcal{D}_k^{g_k}$, if and only if $\bm{d}_1\neq \bm{d}_2$; see Lemma S.1 in Section S2 of the online Supplement. Such property of uniqueness guarantees that, model (\ref{eqnnn5}) can be uniquely transformed into 
\begin{equation}
y(\bm{s}_i)=\sum_{k=1}^{p}\sum_{g_k=1}^{G_k} x_k(\bm{s}_i)\bm{B}(\bm{s}_i)^T \bm{a}^{g_k}_kI(\bm{s}_i\in \mathcal{D}_k^{g_k})+\epsilon(\bm{s}_i).
\label{eqnn6}
\end{equation}

In practice, neither the number of subregions $G_k$ nor the specific subregion $\mathcal{D}_k^{g_k}$ is known. Denote $\bm{a}_{k, i}$ as the  spline coefficient vector for $k$-th covariate at location $\bm{s}_i$, $k=1, \cdots, p$, $i=1, \cdots, n$. From model (\ref{eqnn6}), we know that $\bm{a}_{k, i}$'s are the same for all $\bm{s}_i\in \mathcal{D}_k^{g_k}$.   
To utilize such information, we should encourage homogeneity between spline coefficient vectors, which motivates to minimize the following objective function, 
\begin{equation}
\begin{split}
\frac{1}{2n}\sum_{i=1}^n \{y(\bm{s}_i)&-\sum_{k=1}^{p} x_k(\bm{s}_i)\bm{B}(\bm{s}_i)^T\bm{a}_{k, i}\}^2\\
&+\sum_{k=1}^p\underbrace{\sum_{(i,j)\in \mathbb{E}}P_{\lambda_k}\left(\|\bm{a}_{k, i}-\bm{a}_{k, j}\|_2\right)}_{\text{clustering\ penalty}}+\underbrace{{\sum_{k=1}^p \varrho_k \sum_{i=1}^n \bm{a}_{k, i}^T\PRC\bm{a}_{k, i}}}_{\text{smoothing\ penalty}},
\label{eqn3}
\end{split}
\end{equation} 
where $\|\cdot\|_2$ represents the $L_2$-norm, $P_{\lambda_k}(\cdot)$ is a penalty function for cluster identification, {$\PRC$} is a diagonal matrix determined by the basis functions we choose, and $\{\lambda_k, \varrho_k\}_{k=1}^p$ are tuning parameters determining the strength of penalization. In (\ref{eqn3}),  the smoothing penalty is usually adopted in the context of $P$-splines to address the overparameterized issue because of large $L$, and the  clustering penalty is used to encourage homogeneity for the spline coefficient vectors,  whose corresponding locations are connected by an edge in $\mathbb{E}$. {The edges considered in this paper are undirected, i.e.,  the edge $(i, j)$ equals $(j, i)$.}  The selection of penalty function $P_{\lambda_k}(\cdot)$, the construction of the basis functions $\bm{B}(\cdot)$, the edge set $\mathbb{E}$, and the selection of tuning parameters $\{\lambda_k, \varrho_k\}_{k=1}^p$, are four important ingredients of  (\ref{eqn3}), which are discussed in Section \ref{Implementation}. 
Denote the spline coefficient estimates as $\widehat{\bm{a}}_{k, i}$, then 
$\widehat{\beta}_{k}^{g_k}(\bm{s}_i)=(\widehat{\bm{a}}_{k, i})^T\bm{B}(\bm{s}_i)$. Without causing ambiguity,  we simply denote the procedure of minimizing (\ref{eqn3}) as SCVC.

\vspace{-0.4cm}
\subsection{Implementation details}\label{Implementation}
\vspace{-0.2cm}
\subsubsection{Selection of fused penalty function}\label{penalty-selection}

Among existing literature, LASSO \citep{tibshirani1996regression}, SCAD \citep{fan2001variable} and MCP \citep{zhang2010nearly} are commonly-used penalty functions encouraging sparsity: $\text{LASSO}: P_{\lambda}(t)=\lambda |t|;$ $\text{MCP}: P_{\lambda, \gamma}(t)=(\lambda |t|-\frac{t^2}{2\gamma}) I(|t|\leq \gamma\lambda)+\frac{1}{2}\gamma \lambda^2 I(|t|> \gamma \lambda), \gamma>1$; $\text{SCAD}: P_{\lambda, \gamma}(t)=\lambda |t| I(|t|\leq \lambda)+\frac{2\gamma \lambda |t|-t^2-\lambda^2}{2(\gamma-1)} I(\lambda<|t|<\gamma \lambda)+\frac{\lambda^2 (\gamma+1)}{2} I(|t|\geq \gamma \lambda), \gamma>2$.
 LASSO   assigns large penalties to large values of $t$, thus tends to underestimate $t$, and may not be able to correctly recover the true groups \citep{leng2006note}. To remedy this flaw, SCAD and MCP adopt some concave functions that converge to constants as $t$ increases, which can produce unbiased estimates and are more suitable for identifying the true groups \citep{ma2017concave}. Hence, we adopt these concave penalty functions  in (\ref{eqn3}).

\vspace{-0.3cm}
\subsubsection{Selection of basis functions}\label{basis-function}

The commonly used tensor product spline basis functions are not suitable for spatial data, because the number of its basis functions is huge, which leads to extensive computational burden and numerical instability \citep{crainiceanu2007spatially, opsomer2008non}.

To  address the accompanied issue of tensor product splines, we use low rank radial basis functions \citep{ruppert2003semiparametric}. To be specific, for $\bm{s}=({s}_1, {s}_2)^T\in \mathbb{R}^2$ and the knots $\bm{\kappa}_l\in \mathbb{R}^2,\ l=1, \cdots, L_1$, the low rank radial basis function vector are 
\begin{equation}
1,\ {s}_1, s_2, \ {C(\|\bm{s}-\bm{\kappa}_1\|_2)}, \cdots, {C(\|\bm{s}-\bm{\kappa}_{L_1}\|_{2})},
\label{lowrank}
\end{equation}
where $C(r)$ is a real-valued function; a common choice is $C(r)=r^2 \text{log}\ r$, which corresponds to the thin plate spline.  The number of radial basis functions is $L_1+3$, which is  much smaller than that of the tensor product splines. {To unify the magnitude of the elements in (\ref{lowrank}), similar with  \cite{li2020sparse}, we  normalize the low rank radial basis function vector (\ref{lowrank}), through dividing each element of (\ref{lowrank}) by the mean of its corresponding absolute values calculated over all observed locations.} 
{With the radial basis functions, the corresponding diagonal matrix $\PRC$ in  (\ref{eqn3}) is usually taken to be $\PRC=\text{diag}(\bm{0}_3^T,\bm{1}_{L_1}^T)$}, {where $\bm{0}_3$ is a three-dimensional zero vector, and $\bm{1}_{L_1}$ is a $L_1$-dimensional vector of ones}. Moreover, we take $L_1= \text{max}\{20,  \text{min}(n/4, 40)\}$, see similar choice in \cite{ruppert2003semiparametric}.

The remaining problem is how to select the knots $\bm{\kappa}_l\in \mathbb{R}^2,\ l=1, \cdots, L_1$. In a one-dimensional problem, the knots are usually taken to be equidistant or according to the sample quantiles. However, in the two-dimensional scenario, the equispaced choice tends to waste a lot of knots, and the sample quantile selection does not have a straightforward extension to the two-dimensional space \citep{ruppert2003semiparametric}. Following \cite{ruppert2003semiparametric} and \cite{opsomer2008non}, we select the knots by the \emph{space filling designs} (SFD), in which the knots are closest to the sample locations under the maximal separation principle \citep{johnson1990minimax}. Using the SFD can avoid wasting knots and ensure the coverage of sample locations.  The \pkg{cover.design} function in R package \pkg{Fields} can implement the SFD.

\vspace{-0.3cm}
\subsubsection{Construction of the edge set $\mathbb{E}$} \label{edgMST}
{
Construction of the edge set $\mathbb{E}$ should utilize the spatial neighborhood information of regression coefficients, i.e., coefficients at proximate locations are likely to be similar, possibly resulting from similar conditions for small area \citep{finley2011comparing}. Thus, it is preferable to construct  $\mathbb{E}$, such that only proximate locations are connected, rather than connecting two locations even when they are distant from each other  \citep{ma2017concave}.}

We use the minimum spanning tree (MST) following \cite{li2019spatial}.  Suppose that we have an undirected graph $G=(\mathbb{V}, \mathbb{E}_0)$ with a weight function $d(e)$,  which assigns a weight to each edge $e$ in the edge set $\mathbb{E}_0$, and $\mathbb{V}$ is the set of vertices. 
{In this paper, we take $d(e)$ to be the length of edge $e$ in Euclidean space, $\mathbb{V}$ to be the observed locations, and $\mathbb{E}_0$ to be the edge set by pairwise construction. 
A spanning tree $T=(\mathbb{V}, \mathbb{E})$ is an undirected subgraph of $G$, i.e., $\mathbb{E}\subset \mathbb{E}_0$,  which connects all vertices with no cycles and a minimum number of edges. 
The MST is defined as the spanning tree, whose total edge weight $\sum_{e\in \mathbb{E}} d(e)$ is minimal among all the spanning trees. Thus, MST only connects the proximate locations, utilizing the spatial neighborhood information.} Moreover, MST enjoys two additional advantages. First, it leads to the connectivity of all data points, thus the overfitting issue due to isolated locations, would not happen. {Second, the number of edges in MST is $|\mathbb{V}|-1$ ( $|\mathbb{V}|$ is the number of vertices in $\mathbb{V}$),  which is far less than that of pairwise construction.} Such property allows us to develop an efficient algorithm to minimize (\ref{eqn3}), through utilizing some sparsity structures, see more details in Section \ref{computational-algorithm}. The \pkg{graphminspantree} function in Matlab can be used to find the MST.

\begin{remark}\label{Rremark1}
{
We briefly discuss how the edge set $\mathbb{E}$ constructed by MST influences the cluster identification, and a formal discussion will be presented in Section \ref{asymptotics}. Given $n$ fixed locations, for any specific subregion in (\ref{eqnnn5}), say $\mathcal{D}_1^1$, MST either connects all the locations in $\mathcal{D}_1^1$, resulting in one group, or divides $\mathcal{D}_1^1$ into several groups, where the locations within each group are connected by MST but different groups are not connected. By (\ref{eqn3}), for the former case, all the locations in $\mathcal{D}_1^1$ are expected to be assigned into the same cluster because they are connected; for the latter case, only those connected locations will be assigned into the same cluster, thus $\mathcal{D}_1^1$ will be divided into more than one clusters. 
}
\end{remark}

\vspace{-0.5cm}
\subsubsection{Choices of $\lambda_k$ and $\varrho_k$}

To select the tuning parameters $\{\lambda_k, \varrho_k\}_{k=1}^p$, we adopt  the  Bayesian information criterion (BIC).
{Given $\{\lambda_k, \varrho_k\}_{k=1}^p$, we assume that $\{\widehat{\mathcal{M}}^{g_k^*}_{k}, g_k^*=1, \cdots, \widehat{G}_k^* \}$ are the identified clusters, which is a partition of $\{1, \cdots, n\}$, {and $\widehat{G}_k^*$ is the number of identified clusters}. Within each subgroup, the estimated spline coefficient vectors $\widehat{\bm{a}}_{k, i}$ are equal; see the explicit definition of  $\widehat{\mathcal{M}}^{1}_{k}, \cdots, \widehat{\mathcal{M}}^{\widehat{G}_k^*}_{k}$ in Section \ref{dtcluster}.}
Without loss of generality,  we assume that { $\widehat{\mathcal{M}}^{1}_{k}=\{1, \cdots,  {n_k^1}\}$, $\widehat{\mathcal{M}}^{2}_{k}=\{n_k^1+1, \cdots, n_k^2\}, \cdots, \widehat{\mathcal{M}}^{\widehat{G}_k^*}_{k}=\{{n_k^{\widehat{G}_k^*-1}}+1, \cdots, n\}$,  where $1 \leq {n_k^1}<{n_k^2}<\cdots<{n_k^{\widehat{G}_k^*-1}}<n$, and let $\bm{X}_{\widehat{G^*}}=(\bm{X}_{\widehat{G_1^*}}^1, \cdots, \bm{X}_{\widehat{G_p^*}}^p)$,} where 
	\begin{small}
	\begin{equation*}
	\bm{X}_{\widehat{G}_k^*}^k=\left(
	\begin{array}{ccc}
	x_k(\bm{s}_1)\bm{B}(\bm{s}_1)^T &\cdots&\bm{0}^T  \\[-3mm]
	\vdots &&\vdots\\[-3mm]
	x_k(\bm{s}_{n_k^1})\bm{B}(\bm{s}_{n_k^1})^T &\cdots&\bm{0}^T  \\[-3mm]
	\vdots &\ddots &\vdots  \\[-3mm]
	\bm{0}^T&\cdots &x_k\left(\bm{s}_{ {n_k^{\widehat{G}_k^*-1}}+1}\right)\bm{B}\left(\bm{s}_{ n_k^{\widehat{G}_k^*-1}+1}\right)^T  \\[-3mm]
	\vdots&&\vdots\\[-3mm]
	\bm{0}^T&\cdots &x_k(\bm{s}_{n})\bm{B}(\bm{s}_{n})^T  
	\end{array}
	\right),\  k=1, \cdots, p.
	\end{equation*}
	\end{small}
	Following \cite{tibshirani2012degrees},  the BIC criterion is defined as
	\begin{equation*}
	\text{BIC}(\{\lambda_k, \varrho_k\}_{k=1}^p)=n\ \text{log} \left[\frac{1}{n}\sum_{i=1}^n \Big\{y(\bm{s}_i)-\sum_{k=1}^{p} x_k(\bm{s}_i)\bm{B}(\bm{s}_i)^T\widehat{\bm{a}}_{k, i}\Big\}^2\right]+df *\text{log}\ n,
	\end{equation*}
	where 
	$df=\text{tr}\left\{\bm{X}_{\widehat{G^*}}(\bm{X}_{\widehat{G^*}}^T\bm{X}_{\widehat{G^*}}+2n\widetilde{\PRC})^{-1}\bm{X}_{\widehat{G^*}}^T\right\}$, $\widetilde{\PRC}=\text{diag}(\varrho_1\widetilde{\PRC}_1, \cdots, \varrho_p\widetilde{\PRC}_p)$ and $\widetilde{\PRC}_k=\text{diag}\left\{{n_k^1} \PRC,  ({n_k^2}-{n_k^1})\PRC, \cdots, (n- n_k^{\widehat{G}_k^*-1})\PRC\right\}$, $k=1, \cdots, p$. 

The remaining problem is to  find suitable $\{\lambda_k, \varrho_k\}_{k=1}^p$, which  minimize $\text{BIC}(\{\lambda_k, \varrho_k\}_{k=1}^p)$.   
An intuitive way is to search over a sequence of grid points. However, noticing the number of tuning parameters is greater than one, we need to search a large number of grid points to get a decent result. {To address this, we use the Nelder–Mead method \citep{singer2009nelder}  to minimize $\text{BIC}(\{\lambda_k, \varrho_k\}_{k=1}^p)$, see  
{details in Section S3 of the online Supplement, and a brief description is given as follows.}}
It is a direct search method,  thus is suitable for minimizing $\text{BIC}(\{\lambda_k, \varrho_k\}_{k=1}^p)$  whose derivatives are unknown.
It tries to decrease the function values through a sequence of simplexes, and typically requires only one function evaluation in each iteration step.  Moreover, it can give significant improvements in the first few iterations and quickly produce satisfactory results \citep{singer2009nelder}, because it replaces the worst vertex in the simplex with a better one in each iteration step. According to our experience,  minimizing $\text{BIC}(\{\lambda_k, \varrho_k\}_{k=1}^p)$ with the Nelder–Mead method generally converges within 50 iterations, which means that no more than 50 function evaluations are needed in total, far less than the number of function evaluations using the grid search.

\vspace{-0.3cm}
\subsection{Computational algorithm}\label{computational-algorithm}

Directly minimizing the objective function (\ref{eqn3}) is challenging,  because the penalty function is not separable in $\bm{a}_{k, i}$'s.  We reparameterize (\ref{eqn3}) by introducing a new set of parameters $\bm{\eta}_{k, ij}=\bm{a}_{k, i}-\bm{a}_{k, j}$. Then, minimizing (\ref{eqn3})  is equivalent to 
{
\begin{small}
\begin{eqnarray}
\text{min} \  S(\bm{a},\bm{\eta})&\hspace{-0.6cm}=&\hspace{-0.6cm}\frac{1}{2n}\sum_{i=1}^n \{y(\bm{s}_i)\!-\!\!\sum_{k=1}^{p} x_k(\bm{s}_i)\bm{B}(\bm{s}_i)^T\bm{a}_{k, i}\}^2
\!+\!\sum_{k=1}^p{\sum_{(i,j)\in \mathbb{E}}P_{\lambda_k}\left(\|\bm{\eta}_{k, ij}\|_2\right)}\!+\!{{\sum_{k=1}^p \varrho_k \sum_{i=1}^n 
		\bm{a}_{k, i}^T\PRC\bm{a}_{k, i}}},\nonumber\\
&\!\!\!\!s.t.&\bm{a}_{k, i}-\bm{a}_{k, j}-\bm{\eta}_{k, ij}=0,\quad (i,j)\in \mathbb{E},
\label{au1}
\end{eqnarray} 
\end{small}}
$\!\!\!$where $\bm{a}=(\bm{a}_1^T, \cdots, \bm{a}_p^T)^T$, $\bm{a}_k=(\bm{a}_{k, 1}^T, \cdots, \bm{a}_{k, n}^T)^T$, $k=1, \cdots, p$,  and $\bm{\eta}=(\bm{\eta}_1^T, \cdots, \bm{\eta}_p^T)^T,$ $\bm{\eta}_k=\{\bm{\eta}_{k, ij}^T,\ (i,j)\in \mathbb{E}\}^T$. 
{The above constrained optimization problem can be further converted to an augmented one, 
\begin{eqnarray}
\text{min} \ \ S(\bm{a},\bm{\eta})+\frac{\theta}{2}\sum_{k=1}^p\sum_{(i,j)\in \mathbb{E}} \|\bm{a}_{k, i}-\bm{a}_{k, j}-\bm{\eta}_{k, ij}\|^2_2,\quad
s.t.\ \ \bm{a}_{k, i}-\bm{a}_{k, j}-\bm{\eta}_{k, ij}=0,\ (i,j)\in \mathbb{E},
\label{au}
\end{eqnarray}
where $\theta$ is a positive fixed parameter; see discussion of $\theta$ in Remark \ref{re2}.}
Problems (\ref{au1}) and (\ref{au}) are equivalent, because the quadratic penalty is zero when the constraints are satisfied. To solve the constrained problem (\ref{au}), we use the Lagrangian method, by minimizing
\vspace{-0.2cm}
\begin{small}
\begin{eqnarray}
L(\bm{a},\bm{\eta},\bm{\upsilon})=S(\bm{a},\bm{\eta})
+\frac{\theta}{2}\sum_{k=1}^p\sum_{(i,j)\in \mathbb{E}} \|\bm{a}_{k, i}-\bm{a}_{k, j}-\bm{\eta}_{k, ij}\|^2_2
+\sum_{k=1}^p\sum_{(i,j)\in \mathbb{E}} \bm{\upsilon}_{k, ij}^T\left(\bm{a}_{k, i}-\bm{a}_{k, j}-\bm{\eta}_{k, ij}\right),
\label{eq6}
\end{eqnarray}
\end{small}
$\!\!$where the dual variables $\bm{\upsilon}=(\bm{\upsilon}_1^T, \cdots, \bm{\upsilon}_p^T)^T,$ $\bm{\upsilon}_k=\{\bm{\upsilon}_{k, ij}^T,\ (i,j)\in \mathbb{E}\}^T$, $k=1, \cdots, p$, are the Lagrange multipliers. The expression (\ref{eq6}) is usually called the augmented Lagrangian for (\ref{au1}) \citep{boyd2011distributed}.

We now present the computational algorithm based on ADMM for minimizing  (\ref{eq6}). 
\begin{enumerate}[itemindent=3.5em, label={\emph{Step} \arabic*.}]
	\itemsep=-2pt
	\setcounter{enumi}{-1} 
	\item   Initialize $\bm{\eta}^{(0)}=\bm{0}$ and $\bm{\upsilon}^{(0)}=\bm{0}$.
	\item Given $(\bm{\eta}^{(m)},\bm{\upsilon}^{(m)})$, we update $\bm{a}$ by solving $\partial L(\bm{a},\bm{\eta}^{(m)},\bm{\upsilon}^{(m)})/\partial \bm{a}=\bm{0}$, as
	\begin{equation}
	\begin{split}
	\bm{a}^{(m+1)}=\bm{\Pi}^{-1} 
	\Big[n^{-1}\bm{B}^T\bm{Y}+\sum_{k=1}^p\sum_{(i,j)\in \mathbb{E}}(\bm{e}_{k, i}-\bm{e}_{k, j})\{\theta\bm{\eta}_{k, ij}^{(m)}-\bm{\upsilon}_{k, ij}^{(m)}\}\Big],
	\label{eeq8}
	\end{split}
	\end{equation}
	where $\bm{\Pi}=n^{-1}\bm{B}^T\bm{B}+\theta\{\sum_{k=1}^p\sum_{(i,j)\in \mathbb{E}}(\bm{e}_{k, i}-\bm{e}_{k, j})(\bm{e}_{k, i}-\bm{e}_{k, j})^T\}+2\ols{\PRC}$,  $\bm{Y}=\left(y(\bm{s}_1), \cdots, y(\bm{s}_n)\right)^T$, $\bm{B}=(\bm{B}_1, \cdots, \bm{B}_p)$, $\ols{\PRC}=\text{diag}(\ols{\PRC}_1, \cdots, \ols{\PRC}_p)$,
	\begin{equation*}
	\bm{B}_k=\left(
	\begin{array}{ccc}
	x_k(\bm{s}_1)\bm{B}(\bm{s}_1)^T&\cdots&\bm{0}^T \\[-3mm]
	\vdots&\ddots&\vdots\\[-3mm]
	\bm{0}^T&\cdots&x_k(\bm{s}_n)\bm{B}(\bm{s}_n)^T
	\end{array}
	\right)_{n\times nL},\
	\ols{\PRC}_k=\varrho_k\left(
	\begin{array}{ccc}
	\PRC&\cdots&\bm{0} \\[-3mm]
	\vdots&\ddots&\vdots\\[-3mm]
	\bm{0}&\cdots&\PRC
	\end{array}
	\right)_{nL\times nL},
	\end{equation*}
	$k=1, \cdots, p$, and $\bm{e}_{k, i}$ represents an $npL\times L$ matrix, where the $(g, l)$-th element of $\bm{e}_{k, i}$ is $\bm{e}_{k,i}^{{(g, l)}}$, equal to 1 if $g=l+(k-1)nL+(i-1)L, 1\leq l\leq L$, and 0 otherwise.
	
	\item Given $(\bm{a}^{(m+1)},\bm{\upsilon}^{(m)})$, we update $\bm{\eta}_{k, ij}$ by minimizing
	\begin{equation}
	\frac{1}{2}(\bm{\delta}_{k, ij}^{(m)}-\bm{\eta}_{k, ij})^T(\bm{\delta}_{k, ij}^{(m)}-\bm{\eta}_{k, ij})+\frac{1}{\theta}P_{\lambda_k}\left(\|\bm{\eta}_{k, ij}\|_2\right),
	\label{eq8}
	\end{equation}
	where $\bm{\delta}_{k, ij}^{(m)}=\bm{a}^{(m+1)}_{k, i}-\bm{a}^{(m+1)}_{k, j}+\frac{1}{\theta}\bm{\upsilon}_{k, ij}^{(m)}$. When $P_{\lambda_k}(\cdot)$ is the MCP or SCAD penalty, the minimizer of (\ref{eq8}) has a simple closed-form expression as following:
	
	\begin{enumerate}[label=$\bullet$]
		\itemsep=-3pt
		\item  MCP: $\bm{\eta}_{k, ij}^{(m+1)}=\frac{\gamma}{\gamma-1}M(\bm{\delta}_{k, ij}^{(m)}, \frac{\lambda_1}{\theta})I(\|\bm{\delta}_{k, ij}^{(m)}\|_2\leq \frac{\gamma \lambda_1}{\theta})+\bm{\delta}_{k, ij}^{(m)}I(\|\bm{\delta}_{k, ij}^{(m)}\|_2> \frac{\gamma \lambda_1}{\theta})$, where $M(\bm{z},t)=(1-\frac{t}{\|\bm{z}\|_2})_+ \bm{z}$;
		\item  SCAD: $\bm{\eta}_{k, ij}^{(m+1)}=M(\bm{\delta}_{k, ij}^{(m)}, \frac{\lambda_1}{\theta})I(\|\bm{\delta}_{k, ij}^{(m)}\|_2\leq  \frac{2 \lambda_1}{\theta})+\frac{\gamma-1}{\gamma-2}M(\bm{\delta}_{k, ij}^{(m)}, \frac{\gamma\lambda_1}{(\gamma-1)\theta})I(\frac{2 \lambda_1}{\theta}\leq \|\bm{\delta}_{k, ij}^{(m)}\|_2\leq \frac{\gamma\lambda_1}{\theta})+\bm{\delta}_{k, ij}^{(m)}I(\|\bm{\delta}_{k, ij}^{(m)}\|_2>  \frac{\gamma \lambda_1}{\theta})$.
	\end{enumerate}
	
	\item Update $\bm{\upsilon}_{k, ij}$ as  
	\begin{equation}
	\bm{\upsilon}_{k, ij}^{(m+1)}=\bm{\upsilon}_{k, ij}^{(m)}+\theta\left\{\bm{a}^{(m+1)}_{k, i}-\bm{a}^{(m+1)}_{k, j}-\bm{\eta}_{k, ij}^{(m+1)}\right\}.
	\label{eq11}
	\end{equation}
\end{enumerate}

Repeat Steps 1-3 until a stopping rule is met, and denote the final estimates as $\widehat{\bm{a}},  \widehat{\bm{\eta}}, \widehat{\bm{\upsilon}}$. {In non-convex optimization, it is important to assign appropriate initial values to obtain a good solution. As shown in Step 0, we choose to initialize the ADMM algorithm with $\bm{\eta}^{(0)}=\bm{0}$ and $\bm{\upsilon}^{(0)}=\bm{0}$, which is a common choice \citep{yanglv} and  provides decent results in the simulation studies of Section \ref{simulation-study}.}

The updates from (\ref{eq8})-(\ref{eq11}) are efficient, and the main computational burden concentrates on (\ref{eeq8}), which solves a linear system of equations, i.e.,  $\bm{\Pi}\bm{a}^{(m+1)}=\bm{\zeta}$, where $\bm{\zeta}=n^{-1}\bm{B}^T\bm{Y}+\sum_{k=1}^p\sum_{(i,j)\in \mathbb{E}}(\bm{e}_{k, i}-\bm{e}_{k, j})\{\theta\bm{\eta}_{k, ij}^{(m)}-\bm{\upsilon}_{k, ij}^{(m)}\}$. 
Because $\mathbb{E}$  constructed by MST  contains $n-1$ edges,   the $npL\times npL$ matrix $\bm{\Pi}$ is quite sparse, with proportion of non-zero elements at most $\frac{1}{n}+\frac{5}{npL}$.
Such sparse linear system can be solved efficiently, through storing $\bm{\Pi}$ in a compressed, sparse, column-oriented format, which is implemented by \pkg{sparseMatrix} in R package \pkg{Matrix}. Then, the linear system can be solved efficiently by the function \pkg{solve( $\bm{\Pi}$, $\bm{\zeta}$,   sparse=TRUE)}  in R package \pkg{Matrix}. 

{
When $\mathbb{E}$  is obtained by pairwise construction, the spatial neighborhood information is  not utilized,  {resulting in lots of redundant penalties with the number of $n(n-1)/2$,} which is far larger than $n-1$,  the number of  penalty terms when utilizing the spatial neighborhood information to construct $\mathbb{E}$ through MST.} {It is widely known that,  solving the optimization problem with $n(n-1)/2$ penalties is almost infeasible when $n$ is relatively large, say $n\geq300$. Thus, pairwise construction 
can not be applied to the motivating water mass analysis in Section \ref{real-analysis}, where $n=5130$. 
}

\begin{remark}
	{	Similar with \cite{ma2017concave}, we  track the algorithm based on the primal residual $\bm{R}^{(m+1)}=\{(\bm{a}^{(m+1)}_{k, i}-\bm{a}^{(m+1)}_{k, j}-\bm{\eta}_{k, ij}^{(m+1)})^T,\ (i,j)\in \mathbb{E},\ k=1, \cdots, p\}^T$, and the dual residual $\bm{S}^{(m+1)}=\theta\sum_{k=1}^p\sum_{(i,j)\in \mathbb{E}}(\bm{e}_{k, i}-\bm{e}_{k, j})\{\bm{\eta}_{k, ij}^{(m+1)}-\bm{\eta}_{k, ij}^{(m)}\}$. The algorithm is terminated when $\|\bm{R}^{(m+1)}\|_2\leq \delta_r$ and $\|\bm{S}^{(m+1)}\|_2\leq \delta_s$ for some small positive values $\delta_r$ and $\delta_s$. For a fixed $m$,  according to \cite{boyd2011distributed},  larger $\theta$ usually results in smaller  $\|\bm{R}^{(m+1)}\|_2$ and larger  $\|\bm{S}^{(m+1)}\|_2$. Our numerical experience suggests that, $\theta=1$ is a decent choice,  for which both $\|\bm{R}^{(m+1)}\|_2$ and $\|\bm{S}^{(m+1)}\|_2$ reach small values within a moderate number of iterations. {Such value of $\theta$ is also adopted in \cite{ma2017concave}}.  }
	\label{re2}
\end{remark}

\vspace{-0.9cm}
\section{Asymptotic properties}\label{asymptotics}
\vspace{-0.3cm}
{In this section, we first introduce the definition of true clusters considering the structure of $\mathbb{E}_{\text{MST}}$ (the edge set constructed by MST), and we name these new clusters as ``{\bf{spatial neighborhood true clusters}}'',  {abbreviated as ``{\bf{SpaNeigh true clusters}}"}. We will explain the reason for this name at the beginning of Section 3.1. Then,  we study the oracle property of the SCVC method with $\mathbb{E}_{\text{MST}}$; namely, it works as well as the SpaNeigh true clusters are known. Finally,    we deduce the minimum signal difference requirement for recovering SpaNeigh true clusters.
}

\vspace{-0.5cm}
\subsection{Definition of the SpaNeigh true clusters}
\label{dtcluster}
\vspace{-0.3cm} {
 {
 As discussed in Remark \ref{Rremark1},    the structure of $\mathbb{E}_{\text{MST}}$ plays an important role in cluster identification. Accordingly, we first give the definition of true clusters considering the structure of $\mathbb{E}_{\text{MST}}$, i.e., SpaNeigh true clusters,  and the definition of identified clusters from the SCVC method with $\mathbb{E}_{\text{MST}}$. By the definition in the following, we will see that,   {SpaNeigh true clusters exactly describe the oracle information in Section \ref{sec:intro}, i.e., {for any two locations connected by MST,}  we know whether they belong to the same subregion or not.} Based on the fact that MST only connects proximate locations, SpaNeigh true clusters actually describe the oracle information about whether the location and its neighbors belong to the same subregion or not, and this is the reason for its name.  
To our best knowledge, it is the first time to consider the influence of the structure of $\mathbb{E}_{\text{MST}}$ on cluster identification.
}
}

\begin{figure}[htbp]
	\centering
	\hspace*{-1.8cm}
	\subfigure[]{
		\begin{minipage}[t]{0.6\linewidth}
			\centering
			\includegraphics[width=1.6in,height=1.65in]{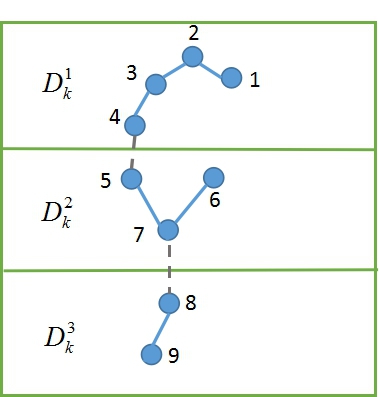}
			\label{ffig1a}
		\end{minipage}%
	}%
	\subfigure[]{
		\begin{minipage}[t]{0.4\linewidth}
			\centering
			\includegraphics[width=1.6in,height=1.6in]{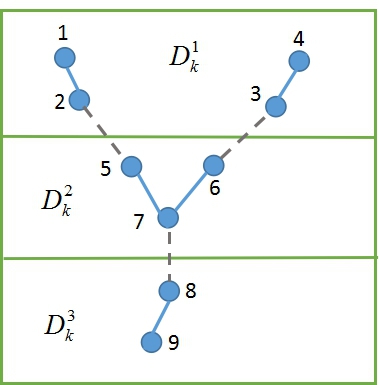}
		\end{minipage}%
		\label{ffig1b}
	}%
	\centering
	\caption{$\mathbb{E}_{\text{MST}}$ consists of within/between-subregion (solid/dashed) connections.}
	\label{Fig1}
\end{figure}

Let $\{\bm{a}_{k, i}^0\}_{i=1}^n$ represent the true values of the spline coefficient vectors, then from model (\ref{eqnn6}), there are $G_k$ distinct values $\bm{a}^{g_k}_k$, {where $\bm{a}_{k, i}^0=\bm{a}^{g_k}_k$ for $\bm{s}_i\in\mathcal{D}_k^{g_k}$, $g_k=1, \cdots, G_k$. Define $\mathcal{G}^{g_k}_k=\{i: \bm{a}_{k, i}^0=\bm{a}^{g_k}_k, 1\leq i\leq n\}$, or equivalently, $\mathcal{G}^{g_k}_k=\{i: \bm{s}_i\in\mathcal{D}_k^{g_k}\}$}, then  $\{\mathcal{G}^{1}_k, \cdots, \mathcal{G}^{G_k}_k\}$ forms a partition of $\{1, \cdots, n\}$, representing the true clusters for $k$-th covariate without considering the structure of $\mathbb{E}_{\text{MST}}$.  
{Considering the structure of $\mathbb{E}_{\text{MST}}$, we give the definition of the SpaNeigh true clusters by the following two steps, where the new cluster is either equal to one of $\mathcal{G}^{g_k}_k$, or a subset, depending on $\mathbb{E}_{\text{MST}}$.}
\begin{enumerate}[]	
\item {For a given $g_k\in\{1, \cdots, G_k\}$, {if for any two  locations $\bm{s}_{i_1}$, $\bm{s}_{i_2}$ in $\mathcal{G}^{g_k}_k$,  i.e., $i_1, i_2 \in \mathcal{G}^{g_k}_k$,} there always exists a path made up of {an edge/some edges} in  $\mathbb{E}_{k}^{g_k}=\{(i, j):(i, j)\in \mathbb{E}_{\text{MST}},\ \text{and}\  i, j\in \mathcal{G}^{g_k}_k\}$,  such that $\bm{s}_{i_1}$ and $\bm{s}_{i_2}$ are connected, then we reserve $\mathcal{G}^{g_k}_k$ as one cluster. Otherwise, { we form a partition of $\mathcal{G}^{g_k}_k$, denoted as $\mathcal{G}^{g_k}_{k, f}, f=1, \cdots, F_k$,} for some positive integer $F_k$, satisfying that for {any two locations} in $\mathcal{G}^{g_k}_{k, f}$,  they are connected through a path, made up of {an edge/some edges} in $\mathbb{E}_{k, f}^{g_k}=\{(i, j):(i, j)\in \mathbb{E}_{\text{MST}},\ \text{and}\  i, j\in \mathcal{G}^{g_k}_{k, f}\}$. { Meanwhile,  for any two locations $\bm{s}_{i_3}\in \mathcal{G}^{g_k}_{k, f}$, $\bm{s}_{i_4}\in \mathcal{G}^{g_k}_{k, f'}$ and $f\neq f'$, {the corresponding edge $(i_3, i_4)\notin \mathbb{E}_{\text{MST}}$;}}} {for example, $\{\bm{s}_1,\bm{s}_2\}$ are not connected with $\{\bm{s}_3,\bm{s}_4\}$ {through $\mathbb{E}_{\text{MST}}$} in Figure \ref{Fig1} (b), so that they are divided into two clusters though they are in the same subregion.}
	
\item Repeating the above step for $g_k=1, \cdots, G_k$,  we obtain either $\mathcal{G}^{g_k}_k$ or $\mathcal{G}^{g_k}_{k, 1},\cdots, \mathcal{G}^{g_k}_{k, F_k}$.  
{We redefine these subgroups as $\{\mathcal{M}^{g_k^*}_{k}\}_{g_k^*=1}^{G^*_k}$, and they are the SpaNeigh true clusters for $k$-th covariate, where $G^*_k\geq G_k$ is the number of new clusters.  A concrete example in the paragraph above Theorem \ref{tthem1} is provided to further illustrate  $\{\mathcal{M}^{g_k^*}_{k}\}_{g_k^*=1}^{G^*_k}$.}
\end{enumerate}

For the identified clusters, we assume that there are $\widehat{G}_k$ distinct values in $\{\widehat{\bm{a}}_{k, i}\}_{i=1}^n$, denoted as $\widehat{\bm{a}}^{g_k}_k, g_k=1, \cdots, \widehat{G}_k$.  {As a counterpart of the SpaNeigh true clusters}, the identified clusters from the SCVC method with $\mathbb{E}_{\text{MST}}$,  are obtained from the above two steps by replacing $\bm{a}_{k, i}^0$ with $\widehat{\bm{a}}_{k, i}$,  ${\bm{a}}^{g_k}_k$ with $\widehat{\bm{a}}^{g_k}_k$, and $G_k$ with $\widehat{G}_k$.  We denote them as $\{\widehat{\mathcal{M}}^{1}_{k}, \cdots, \widehat{\mathcal{M}}^{\widehat{G}_k^*}_{k}\}$, {where $\widehat{G}_k^*\geq \widehat{G}_k$ is the number of identified clusters from the SCVC method with $\mathbb{E}_{\text{MST}}$.}

By definition,  {the SpaNeigh true cluster} $\mathcal{M}^{g_k^*}_{k}$ is a subset of   $\mathcal{G}^{g_k}_k$ for some $g_k\in\{1, \cdots, G_k\}$,  $g_k^*=1, \cdots, G^*_k$,  and {$\{\mathcal{M}^{1}_{k}, \cdots, \mathcal{M}^{G_k^*}_{k}\}$ exactly describes the oracle information in Section \ref{sec:intro}, i.e., {for any two locations connected by MST,} we know whether they belong to the same subregion  or not. Based on the fact that MST only connects proximate locations,  $\{\mathcal{M}^{1}_{k}, \cdots, \mathcal{M}^{G_k^*}_{k}\}$  actually describes the oracle information about whether the location and its neighbors belong to the same subregion or not.  However, $\{\mathcal{G}^{1}_k, \cdots, \mathcal{G}^{G_k}_k\}$ describes the oracle information,  that for {any two locations} even they are distant from each other, we know whether  they belong to the same subregion or not. Under the framework of spatial data, people may not care whether  {two locations} belong to the same subregion if they are distant from each other, and just want to know whether the location and its neighbors can be assigned into the same cluster. From that point of view, people may only need to recover  $\{\mathcal{M}^{1}_{k}, \cdots, \mathcal{M}^{G_k^*}_{k}\}$ instead of $\{\mathcal{G}^{1}_k, \cdots, \mathcal{G}^{G_k}_k\}$. 
Moreover, we use a concrete example for further illustration, see Figure {\ref{Fig1}}.
{In Figure \ref{ffig1a}, the locations within subregion $\mathcal{D}_k^1$ (or $\mathcal{D}_k^2$, or $\mathcal{D}_k^3$) are relatively close to each other, so that all of them are connected by MST.  Then, we have 
	 $G_k=G^*_k=3$ and $\{\mathcal{G}^{1}_k, \cdots, \mathcal{G}^{G_k}_k\}$ $=\{\mathcal{M}^{1}_{k}, \cdots, \mathcal{M}^{G_k^*}_{k}\}=\left\{\{1, 2, 3, 4\}, \{5, 6, 7\}, \{8, 9\}\right\}$.
In Figure \ref{ffig1b}, within subregion $\mathcal{D}_k^1$,  the location set $\{1, 2\}$  is distant from  the location set $\{3, 4\}$, so that MST does not connect them. Then, we have $G^*_k=4$ with $\{\mathcal{M}^{1}_{k}, \cdots, \mathcal{M}^{G_k^*}_{k}\}=\left\{\{1, 2\}, \{3, 4\}, \{5, 6, 7\}, \{8, 9\}\right\}$, and 
$G_k=3$ with $\{\mathcal{G}^{1}_k, \cdots, \mathcal{G}^{G_k}_k\}=\left\{\{1, 2, 3, 4\},  \{5, 6, 7\}, \{8, 9\}\right\}$.  For the former situation where $\{\mathcal{G}^{1}_k, \cdots, \mathcal{G}^{G_k}_k\}=\{\mathcal{M}^{1}_{k}, \cdots, \mathcal{M}^{G_k^*}_{k}\}$, we call it ``{\bf{MST-equal}}'';  for the latter situation where $\{\mathcal{G}^{1}_k, \cdots, \mathcal{G}^{G_k}_k\}\neq\{\mathcal{M}^{1}_{k}, \cdots, \mathcal{M}^{G_k^*}_{k}\}$, we call it ``{\bf{MST-unequal}}''.
}}

\begin{Theorem}
	Under  $\mathbb{E}_{\emph{MST}}$, the sets of SpaNeigh true clusters $\{\mathcal{M}^{1}_{k}, \cdots, \mathcal{M}^{G_k^*}_{k}\}$ and identified clusters $\{\widehat{\mathcal{M}}^{1}_{k}, \cdots, \widehat{\mathcal{M}}^{\widehat{G}_k^*}_{k}\}$, are existing and unique, $k=1, \cdots, p$.
	\label{tthem1}
\end{Theorem}

{
\begin{remark}
\label{remmark3}
As discussed in Section \ref{edgMST}, MST is influenced by the definition of distance, and so is $\{\mathcal{M}^{1}_{k}, \cdots, \mathcal{M}^{G_k^*}_{k}\}$. {Different  distance metrics essentially reflect different beliefs on the spatial neighborhood information, because they determine the similarity between locations.} For instance, two locations, which are close to each other under Euclidean distance, can be distant from each other under other distances, such as, geodesic distance, see \cite{wang2007low}.
Thus, for a specific problem,  a proper distance leads to proper use of spatial neighborhood information, thus the corresponding  $\{\mathcal{M}^{1}_{k}, \cdots, \mathcal{M}^{G_k^*}_{k}\}$ based on such distance may be more reasonable. 
Euclidean distance adopted in this paper is widely used in spatial analysis, when the shape of a domain is regular. When the shape of a domain  is irregular with complex boundaries or interior gaps and holes, geodesic distance, that is, the length of the shortest path within the domain between two points,  may more accurately reflect the spatial neighborhood information than Euclidean distance, as it considers the complex shape of the domain, see details in \cite{wang2007low}.   
\end{remark}
}

\vspace{-0.7cm}

\subsection{Oracle properties of the SCVC method with $\mathbb{E}_{\text{MST}}$}
\vspace{-0.3cm}
When {the SpaNeigh true clusters}, i.e., $\mathcal{M}^{1}_{k}, \cdots, \mathcal{M}^{G_k^*}_{k}$, $k=1, \cdots, p$,  are known, the oracle estimator for  $\bm{a}=(\bm{a}_1^T, \cdots, \bm{a}_p^T)^T$ is
	\begin{equation}
	\widehat{\bm{a}}^{\text{or}}=\underset{ \bm{a}_k\in \mathcal{N}_{\mathcal{G}}^k,\, k=1, \cdots, p}{\text{arg}\  \text{min}} \frac{1}{2n}\sum_{i=1}^n \{y(\bm{s}_i)-\sum_{k=1}^{p} x_k(\bm{s}_i)\bm{B}^T(\bm{s}_i)\bm{a}_{k, i}\}^2+{{\sum_{k=1}^p\varrho_k\sum_{i=1}^n
			\bm{a}_{k, i}^T\PRC\bm{a}_{k, i}}},
	\label{eqn13}
	\end{equation}
	where $\mathcal{N}_{\mathcal{G}}^k$ is the subspace of $\mathbb{R}^{nL}$, defined as
	\begin{equation*}
	\mathcal{N}_{\mathcal{G}}^k=\{\bm{a}_k=(\bm{a}_{k, 1}^T, \cdots, \bm{a}_{k, n}^T)^T\in \mathbb{R}^{nL}: \bm{a}_{k, i}=\bm{a}_{k, j}~\text{for any}\  i, j\in \mathcal{M}^{g_k^*}_{k}, 1\leq g_k^*\leq G_k^* \}. 
	\end{equation*}

Let $\bm{a}^0$ be the true value of $\bm{a}$,  $\widetilde{\varrho}=\underset{1\leq k\leq p}{\text{max}}\ \varrho_k$, $|\mathcal{M}_\text{min}|=\underset{1\leq k\leq p, 1\leq g_k^*\leq G_k^*}{\text{min}} |\mathcal{M}^{g_k^*}_{k}|$ and $|\mathcal{M}_\text{max}|=\underset{1\leq k\leq p, 1\leq g_k^*\leq G_k^*}{\text{max}} |\mathcal{M}^{g_k^*}_{k}|$, where $ |\mathcal{M}^{g_k^*}_{k}|$ is the number of elements in $ \mathcal{M}^{g_k^*}_{k}$.  For any numbers $a_n>0$ and $b_n>0$, let $a_n\asymp b_n$ represent $\text{lim}_{n\to \infty} a_n/b_n=c$ for some $c>0$, and $a_n>>b_n$ represent $a_n^{-1}b_n=o(1)$. For any $s\times t$ matrix $\bm{A}=(A_{ij})_{i=1, j=1}^{s, t}$, denote $\|\bm{A}\|_\infty=\underset{1\leq i\leq s}{\text{max}} \sum_{j=1}^t |A_{ij}|$.

\setcounter{theorem}{1}
\begin{theorem}
	Under the Assumptions (A1)-(A6) in the online Supplement, if  
	$|\mathcal{M}_\emph{min}|$ $>> \sqrt{ \sum_{k=1}^p G_k^*} \sqrt{n\log n}$ and $\widetilde{\varrho}<< {\sqrt{\log n}}/({\sqrt{n}|\mathcal{M}_\emph{max}|\, \|{\bm{a}}^0\|_\infty})$, we have
	\begin{equation*}
	\|\widehat{\bm{a}}^\emph{or}-\bm{a}^0\|_{\infty}\leq r_n,
	\end{equation*}
	with probability approaching one, where $r_n\asymp\sqrt{ \sum_{k=1}^p G_k^*} \sqrt{n\log n}/|\mathcal{M}_\emph{min}|$.
	\label{them2}
\end{theorem}

\begin{remark}
Let $G_{\emph{max}}^*=\underset{1\leq k\leq p}{\emph{max}}\ G_k^*$, 	by the condition $|\mathcal{M}_\emph{min}|>> \sqrt{ \sum_{k=1}^p G_k^*} \sqrt{n\log n}$ in Theorem \ref{them2}, we have ${G_{\emph{max}}^*}^{3/2} \sqrt{n\log n}<<n$, thus the maximum number of true clusters need to satisfy $G_{\emph{max}}^*<<(n/\log n)^{1/3}$.
\end{remark}

{Now we consider the theoretical properties of the SCVC method with $\mathbb{E}_{\text{MST}}$, in terms of cluster identification and coefficient estimation.} It is expected that,  the signal difference between different clusters   plays an important role, which is measured by the difference of the true spline coefficient vectors  in different groups, and larger signal difference makes it easier for true cluster recovering. For $k$-th covariate,  we define the minimum signal difference under the structure of $\mathbb{E}_{\text{MST}}$ as
\begin{equation}
\vartheta_{k}= \underset{
	\begin{scriptsize}
	\begin{split}
	\vspace{-1cm}	 &i\in \mathcal{M}^{g}_{k}, j\in\mathcal{M}^{g'}_{k}, g\neq g',\\
	&\quad\quad\quad (i, j)\in \mathbb{E}_{\text{MST}}.
	\end{split}
	\end{scriptsize}
}{\text{min}} \|\bm{a}_{k, i}^0-\bm{a}_{k, j}^0 \|_2, \quad k=1, \cdots, p.
\label{eqn15}
\end{equation}
Therefore, to make $\vartheta_{k}>0$, we only require that the true spline coefficient vector pairs $\bm{a}_{k, i}^0$ and $\bm{a}_{k, j}^0$ are different, if the corresponding locations $\bm{s}_i$ and $\bm{s}_j$ belong to two proximate different subregions. 
{Take the case in Figure \ref{ffig1b} as an example,  we have $\vartheta_{k}=\underset{(i, j)\in \{(2, 5), (7, 8)\}}{\text{min}}$ $ \|\bm{a}_{k, i}^0-\bm{a}_{k, j}^0 \|_2$. } More discussion about $\vartheta_{k}$ can be found in  Remark \ref{re6}.

\begin{theorem}\label{them3}
	Suppose that the assumptions in Theorem \ref{them2} hold, $\widetilde{\varrho} << {\sqrt{\log n}}/\{n(\|{\bm{a}}^0\|_\infty$ $+r_n)\}$, $\vartheta_{k}>>\lambda_k$ and 
	\begin{equation*}
	\lambda_k>> \underset{  1\leq g_k^*\leq G_k^*}{\emph{max}}\{r_n, |\mathcal{M}^{{g_k^*}}_{k}| \sqrt{\log n}/n\}, \quad k=1, \cdots, p. 
	\end{equation*}
Then, there exists a local minimizer $ \widehat{\bm{a}}$ of the objective function (\ref{eqn3}) with  $\mathbb{E}_{\emph{MST}}$,  satisfying
	\begin{equation*}
	P\left( \widehat{\bm{a}}= \widehat{\bm{a}}^\emph{or}\right)\to 1.
	\end{equation*}
\end{theorem}
Theorem \ref{them3} shows that,  the oracle estimator $ \widehat{\bm{a}}^\text{or}$ is a local minimizer of the objective function (\ref{eqn3}) with  $\mathbb{E}_{\text{MST}}$,  with probability approaching one. By the definition of  $\{\widehat{\mathcal{M}}^{1}_{k}, \cdots, \widehat{\mathcal{M}}^{\widehat{G}_k^*}_{k}\}$, we have the following corollary.

\begin{cor}\label{cor1}
	Suppose the assumptions in Theorem \ref{them3}  $\,$ hold, we have
	\begin{equation*}
	P\left(\{\widehat{\mathcal{M}}^{1}_{k}, \cdots, \widehat{\mathcal{M}}^{\widehat{G}_k^*}_{k}\}=\{{\mathcal{M}}^{1}_{k}, \cdots, {\mathcal{M}}^{{G_k^*}}_{k}\}\right)\to 1,~k=1, \cdots, p.
	\end{equation*}	
\end{cor}
Corollary \ref{cor1} shows that,
the SCVC method with $\mathbb{E}_{\text{MST}}$ can identify the SpaNeigh true clusters $\{{\mathcal{M}}^{1}_{k}, \cdots, {\mathcal{M}}^{{G_k^*}}_{k}\}$ with probability approaching one.

\begin{remark}

	We discussed the minimum signal difference $\vartheta_{k}$ in Theorem \ref{them3}.  From the conditions $\vartheta_{k}>>\lambda_k$ and 
	$\lambda_k>>\underset{  1\leq g_k^*\leq G_k^*}{\emph{max}}\{r_n, |\mathcal{M}^{{g_k^*}}_{k}| \sqrt{\log n}/n\},$ we have 
	\begin{equation*}
	\vartheta_{k}>>\underset{  1\leq g_k^*\leq G_k^*}{\emph{max}}\{r_n, |\mathcal{M}^{{g_k^*}}_{k}| \sqrt{\log n}/n\}, \quad k=1, \cdots, p.
	\end{equation*}
	Suppose $|\mathcal{M}_\emph{min}|\asymp |\mathcal{M}_\emph{max}|$, 
	\begin{enumerate}[label={(\roman*)}]
		\itemsep=-3pt
		\item when ${n^{1/5}}<<G_k^*<<n^{1/3} (\log n)^{-1/3}$, then  $r_n>>|\mathcal{M}^{{g_k^*}}_{k}| \sqrt{\log n}/n$,   we have the minimum signal difference satisfying $\vartheta_{k}>>r_n$;
		\item when $0<G_k^*<<n^{1/5}$, then  $r_n<<|\mathcal{M}^{{g_k^*}}_{k}| \sqrt{\log n}/n$,\quad   so we need $\vartheta_{k}>>$ $|\mathcal{M}^{{g_k^*}}_{k}| \sqrt{\log n}/n$, and if  the number of clusters is fixed, we have $\vartheta_{k}>>\sqrt{\log n}$, see a similar rate in \cite{ke2015homogeneity}.
	\end{enumerate}
	\label{re6}
\end{remark}
\vspace{-1.5cm}

\section{Simulation studies}\label{simulation-study}
\vspace{-0.4cm}
We present two simulation studies to illustrate the finite-sample performance of SCVC method, based on the SCAD penalty with $\gamma=3.7$ \citep{fan2001variable, zou2008one}. 
Results based on MCP are similar and thus omitted. In the first study, we consider constant clustered coefficients, i.e., the coefficients remain constant within each cluster. In the second study, the true coefficients are clustered,  varying smoothly within each cluster. 

{
	In each study,   we  consider  two different spatially clustered patterns
	in the square domain $[0,1]\times[0,1]$, made up of 1000 locations.}
Based on these locations,  we generate the covariates as $x_1(\bm{s}_i)\equiv 1$ and $\{x_2(\bm{s}_i)\}_{i=1}^{1000}$ as realizations of a spatial Gaussian process with mean zero and covariance function $\text{Cov}\{x(\bm{s}_i),x(\bm{s}_j)\}=\text{exp}(-\|\bm{s}_i-\bm{s}_j\|/\phi)$, where $\phi$ is the range parameter, and   $\phi=0.1$,  $1$ correspond to  weak and strong spatial correlations.  Based on these  locations and covariates,  the data generating process is
\begin{equation*}
y(\bm{s}_i)=x_1(\bm{s}_i)\beta_1(\bm{s}_i)+x_2(\bm{s}_i)\beta_2(\bm{s}_i)+\epsilon(\bm{s}_i),\ i=1, \cdots, 1000,
\end{equation*}
where  $\epsilon(\bm{s}_i)\overset{i.i.d.}{\sim}N(0,0.1^2)$, and we run 100 replicates to examine the behavior of SCVC in parameter estimation and cluster identification.

For comparison, we  include the  SCC \citep{li2019spatial},  the GWR \citep{fotheringham2003geographically} method, {and the common $P$-spline estimator (PSE) which is obtained by assuming the spline coefficient vectors in (\ref{eqn3}) are the same for any locations.}  The SCC can deal with the spatially clustered pattern with constant regression coefficients within each subregion, while the GWR and the PSE can deal with the scenario when the regression coefficients vary smoothly over the whole region. {Moreover, we also include the method by replacing the LASSO penalty in SCC with the SCAD penalty, and name it as ``SCC*''. The algorithm and code provided by \cite{li2019spatial} can not deal with SCC*, and  we modify the ADMM algorithm in Section \ref{computational-algorithm} to implement SCC*. Because the optimization problem for SCC* is non-convex, we try many types of initial values, such as zero, random initial values and initial values from the SCC estimates,  and find that the initial values from the SCC estimates result in the best performance of SCC* among these types of initial values.  Thus, for comparison purpose,  
 we report the results of SCC*   with the initial values from the SCC estimates. }
To quantify the performance of each method, we consider three  criteria. (i) $\text{MSE}_{\beta_k}$: the mean-squared error   for $k$-th covariate, defined as
\begin{equation*}
\text{MSE}_{\beta_k}=\frac{1}{n}\sum_{i=1}^n \{\widehat{\beta}_k(\bm{s}_i)-{\beta}_k(\bm{s}_i)\}^2, \quad k=1, 2.
\end{equation*}
(ii) $\text{RI}_k$: the rand index for $k$-th covariate, a commonly used criterion in clustering analysis, which measures the percentage of correct identifications, defined as
\begin{equation*}
\text{RI}_k=\frac{\text{TP}_k+\text{TN}_k}{\text{TP}_k+\text{FP}_k+\text{FN}_k+\text{TN}_k},\quad k=1, 2.
\end{equation*}
{$\text{TP}_k/\text{FP}_k$ (true positive/false positive) is the number of location pairs from different subregions assigned to different clusters/the same cluster; $\text{TN}_k/\text{FN}_k$ (true negative/false negative) is the number of pairs from the same subregion  assigned to the same cluster/different clusters.} Higher values of $\text{RI}_k$ indicate better agreement of the identified clusters with the true subregions. (iii) $\text{IC}_k$: the number of identified clusters for $k$-th covariate. All these criteria are averaged over 100 replicates.

The tuning parameters in SCVC, SCC, SCC*, and PSE are chosen by BIC, and for the GWR, we employ an exponential kernel function with optimal bandwidth chosen by the cross-validation. The SCC and GWR are realized by the code from \cite{li2019spatial},  SCC*  by modifying the  ADMM algorithm in Section \ref{computational-algorithm}, {and PSE by its closed expression.}

\vspace{-0.6cm}
\subsection{Study 1: Constant clustered coefficients}
\begin{figure}[htbp]
	\centering	
	\vspace{-1cm}
	
	\subfigure[Constant clustered coefficients  ]{
		\begin{minipage}[t]{0.8\linewidth}
			\setlength{\abovecaptionskip}{-2cm} 
			\centering
			\hspace*{-4cm}\includegraphics[width=8in,height=2.6in]{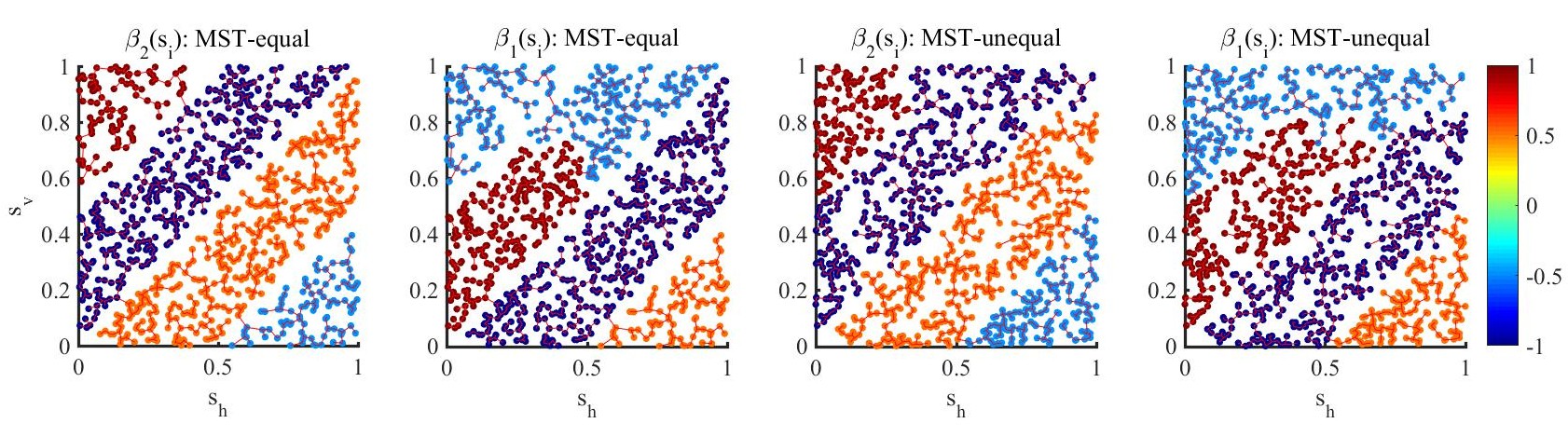}
			\label{ffigure4a}
		\end{minipage}%
	}%
	
	\subfigure[Smooth-varying clustered coefficients]{
		\begin{minipage}[t]{0.8\linewidth}
			\setlength{\abovecaptionskip}{-2cm} 
			\centering
			\hspace*{-4cm}\includegraphics[width=8in,height=2.6in]{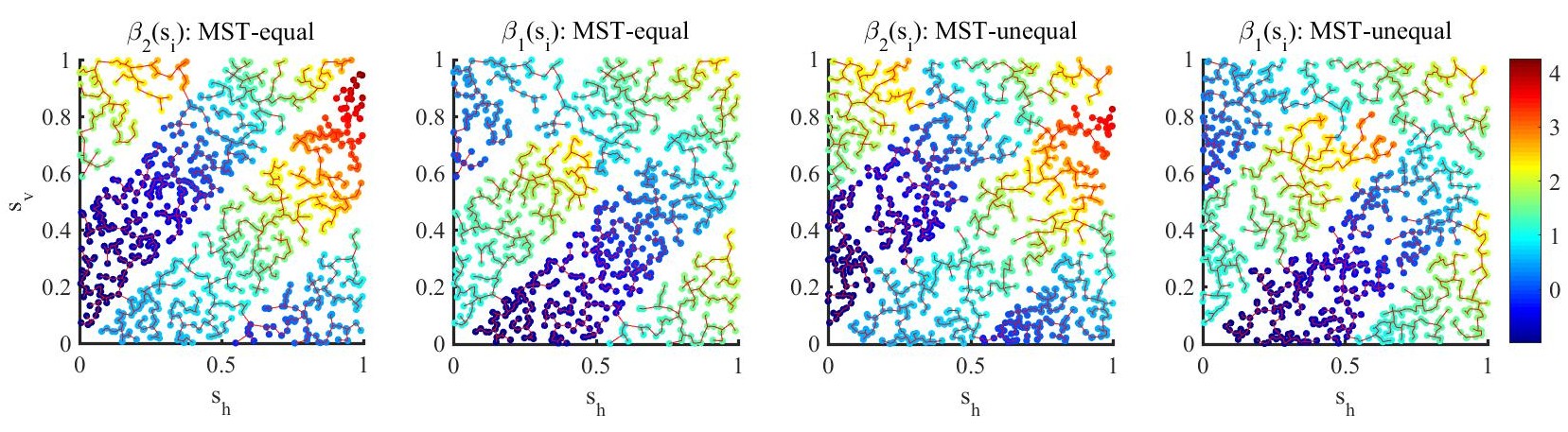}
			\label{ffigure4b}
		\end{minipage}%
	}%
	\centering
	\caption{{The points/colors represent the locations/coefficient values, and the solid lines represent the edges in MST. From the top left cluster to the bottom right one: in (a), $\beta_2(\bm{s})$ equals 1, -1, 0.5, -0.5, respectively, and $\beta_1(\bm{s})$ equals -0.5, 1, -1, 0.5, respectively; in (b), with $\bm{s}=(s_h, s_v)$ and $s_{hv}=s_h+s_v$, $\beta_2(\bm{s})$ equals $1+s_{hv}^2$, $-1+s_{hv}^2$, $0.5+s_{hv}^2$, $-0.5+s_{hv}^2$, respectively, and $\beta_1(\bm{s})$ equals $-0.5+s_{hv}^{1.5}$, $1+s_{hv}^2$, $-1+s_{hv}^{1.7}$, $0.5+s_{hv}^{1.5}$, respectively.}}
	\label{ffigure4}
\end{figure}

\vspace{-0.3cm}
The true regression coefficients $\{\beta_k(\bm{s}_i)\}_{k=1}^2$ in this study are spatially clustered and remain constant within each cluster, i.e., the SCC model is the true model. Moreover, two different spatially clustered patterns are considered, as shown in Figure \ref{ffigure4a}.   {The spatially clustered patterns in the left two panels of Figure \ref{ffigure4a}} represent {{MST-equal}},  i.e.,  $\{\mathcal{G}^{1}_k, \cdots, \mathcal{G}^{G_k}_k\}=\{\mathcal{M}^{1}_{k}, \cdots, \mathcal{M}^{G_k^*}_{k}\}$, $k=1, 2$;  {the spatially clustered patterns in the right two panels} of Figure \ref{ffigure4a} represent {{MST-unequal}}, i.e.,  $\{\mathcal{G}^{1}_k, \cdots, \mathcal{G}^{G_k}_k\}$ $\neq \{\mathcal{M}^{1}_{k}, \cdots, \mathcal{M}^{G_k^*}_{k}\}$, $k=1, 2$.
  Details about generating these two different clustered patterns in Figure \ref{ffigure4} can be found in Section S4 of the online Supplement.

{
Table \ref{table3} compares the MSE, RI, and IC for SCVC, SCC,  GWR,  SCC*, and PSE, with patterns specified either by  MST-equal or MST-unequal in Figures \ref{ffigure4a}.
The results in Table \ref{table3} suggest that the SCVC generally outperforms the SCC,  GWR, and PSE. First,  the MSEs of SCVC are smaller than those of SCC,  GWR, and PSE. Second, SCVC gives a more reasonable number of identified clusters than SCC. Third, SCVC results in higher RI values than SCC. 
One possible explanation is that,  compared to the SCAD penalty in the SCVC, the LASSO penalty in the SCC  tends to result in a larger number of identified clusters and lower efficiency for estimation.  This explanation is supported by the results of SCC*, which generally show better performance than SCC due to utilizing SCAD penalty. Now we compare the results of SCVC and SCC*, where both methods utilize SCAD penalty. For cluster identification, both methods produce similar results. For coefficient estimation, SCVC performs better than SCC for the estimation of $\beta_1(\bm{s}_i)$ under MST-equal (or MST-unequal) with weak correlation; for the rest cases, SCVC and SCC* perform similarly or SCC* performs} {slightly better.} {One possible reason for the underperformance of SCC* in some cases is the sensitivity of SCC* to the initial value. If setting the initial values as the true values, 
the {\it infeasible} SCC* uniformly outperforms SCVC for coefficient estimation and cluster identification,  see Section S5 in the online Supplement}.

\begin{table}[htbp] 
	\vspace{-1.5cm}
   \scriptsize
    \renewcommand\arraystretch{0.8}
	\setlength{\abovecaptionskip}{0pt}
	\setlength{\belowcaptionskip}{10pt}
	\begin{threeparttable}[b]
		\caption{Summary of Study 1, with constant coefficient within each cluster.} 
		\begin{tabular}{p{1.8cm}<{\centering}p{1.5cm}<{\centering}p{1.5cm}<{\centering}p{1.5cm}<{\centering}p{1.5cm}<{\centering}p{1.3cm}<{\centering}p{1.3cm}<{\centering}p{1cm}<{\centering}p{1cm}<{\centering}} 
\hline
\hline
\multicolumn{1}{p{1.8cm}<{\centering}}{ Pattern}&\multicolumn{1}{p{1.5cm}<{\centering}}{Correlation} & \multicolumn{1}{p{1.5cm}<{\centering}}{Methods}& \multicolumn{1}{p{1.5cm}<{\centering}}{$\text{MSE}_{\beta_2}$}&\multicolumn{1}{p{1.5cm}<{\centering}}{$\text{MSE}_{\beta_1}$}&\multicolumn{1}{p{1.25cm}<{\centering}}{$\text{RI}_2$}&\multicolumn{1}{p{1.25cm}<{\centering}}{$\text{RI}_1$}&\multicolumn{1}{p{1.25cm}<{\centering}}{$\text{IC}_2$} &\multicolumn{1}{p{1.25cm}<{\centering}}{$\text{IC}_1$}\\
			\cmidrule(lr){4-5}\cmidrule(lr){6-7}\cmidrule(lr){8-9}
			&&SCVC&0.021&0.010&99.72&99.96&5.84&4.48\\
			&&&(0.001)&(0.001)&(0.02)&(0.01)&(0.04)&(0.05)\\
			&&SCC&0.029&0.079&86.04&78.69&20.65&20.51\\
		    &&&(0.001)&(0.001)&(0.43)&(0.25)&(0.33)&(0.34)\\
		    &weak&SCC*&0.024&0.090&99.37&99.60&7.00&4.00\\
		    &&&(0.001)&(0.001)&(0.00)&(0.00)&(0.00)&(0.00)\\
			&&GWR&0.199&0.214&-&-&-&-\\
			&&&(0.004)&(0.004)&-&-&-&-\\
			&&PSE&0.855&0.907&-&-&-&-\\
			MST-equal&&&(0.002)&(0.003)&-&-&-&-\\
			\cmidrule(lr){3-9}
			&&SCVC&0.086&0.099&99.58&98.46&4.49&4.19\\
			&&&(0.006)&(0.013)&(0.15)&(0.29)&(0.06)&(0.04)\\
			&&SCC&0.197&0.288&75.15&73.26&45.03&39.19\\
			&&&(0.004)&(0.005)&(0.14)&(0.17)&(0.59)&(0.39)\\
			&strong&SCC*&0.050&0.130&99.14&99.05&8.00&8.00\\
			&&&(0.001)&(0.001)&(0.00)&(0.00)&(0.00)&(0.00)\\
			&&GWR&1.364&1.932&-&-&-&-\\
			&&&(0.023)&(0.043)&-&-&-&-\\
			&&PSE&1.741&1.505&-&-&-&-\\
			&&&(0.009)&(0.008)&-&-&-&-\\
			\hline
		    &&SCVC&0.096&0.054&85.51&87.77&8.26&12.78\\
		    &&&(0.002)&(0.003)&(0.01)&(0.02)&(0.17)&(0.07)\\
		    &&SCC&0.118&0.200&78.59&78.59&49.03&42.35\\
		    &&&(0.001)&(0.002)&(0.13)&(0.13)&(0.64)&(0.55)\\
		    &weak&SCC*&0.130&0.155&85.11&87.72&11.00&12.82\\
		    &&&(0.001)&(0.001)&(0.00)&(0.00)&(0.00)&(0.04)\\
		    &&GWR&0.447&0.707&-&-&-&-\\
		    MST-unequal&&&(0.003)&(0.004)&-&-&-&-\\
		    &&PSE&1.088&1.386&-&-&-&-\\
		    &&&(0.002)&(0.003)&-&-&-&-\\
		    \cmidrule(lr){3-9}
		    &&SCVC&0.850&2.307&81.22&82.66&6.78&8.79\\
		    &&&(0.024)&(0.075)&(0.15)&(0.15)&(0.10)&(0.25)\\
		    &&SCC&1.321&4.067&77.60&75.86&67.76&40.22\\
		    &&&(0.018)&(0.057)&(0.14)&(0.07)&(1.24)&(0.79)\\
		    &strong&SCC*&0.926&2.732&81.37&84.06&26.26&14.69\\
		    &&&(0.004)&(0.001)&(0.03)&(0.03)&(0.10)&(0.07)\\
		    &&GWR&4.450&3.024&-&-&-&-\\
		    &&&(0.137)&(0.043)&-&-&-&-\\
		    &&PSE&2.494&2.715&-&-&-&-\\
		    &&&(0.008)&(0.018)&-&-&-&-\\
			\hline
			\hline
		\end{tabular}
		\label{table3}
		\begin{tablenotes} 
			\item SCVC: spatially clustered varying coefficient method; SCC: spatially clustered coefficient regression based on LASSO; SCC*: spatially clustered coefficient regression based on SCAD; GWR: geographically weighted regression; PSE: $P$-spline estimator. $\text{MSE}_{\beta_k}$/$\text{RI}_k$/$\text{IC}_k$: mean squared error $(\times 10)$/rand index $(\times 100)$/number of identified clusters, for $k$-th covariate, $k=1, 2$. Values in the parentheses are the standard errors. Note that GWR and PSE can not identify clusters.
		\end{tablenotes}
	\end{threeparttable}
\end{table}

\vspace{-0.5cm}

\subsection{Study 2: Smooth-varying clustered coefficients}\label{Smooth-varying}
\vspace{-0.3cm}
In this study, the sample locations and clusters are the same as those in Study 1, except that the coefficients within each cluster are smooth-varying, see Figure \ref{ffigure4b}. {The left two and right two panels in Figure \ref{ffigure4b} represent MST-equal and  MST-unequal, respectively. }

Table \ref{table1} summarizes the comparison of the five methods under two different spatially clustered patterns, i.e.,  MST-equal and MST-unequal. 
In this scenario, the assumptions for SCC, SCC*,  GWR, and PSE are violated, thus SCVC performs much better.
First, SCVC clearly outperforms SCC, SCC*,  GWR, and PSE for coefficient estimation, with considerably smaller MSE in all settings. 
Second, SCVC yields a reasonable number of clusters, while SCC and SCC* lead to a much larger number of clusters. The main reason is that,   the SCC and SCC* require the coefficients to be constant within each cluster,
thus lead to more clusters when the true coefficients are varying.
Last, SCVC results in much higher RI values than SCC and SCC*, suggesting better agreement of the clusters. 

 {Additional simulations with the true coefficients smooth over the whole region, i.e., the assumption made in GWR and PSE holds,  are provided in Section S6 of the online Supplement. The corresponding results show  PSE performs slightly better than SCVC, and SCVC performs better than other remaining methods, see detailed discussion there.}

\begin{table}[htbp] 
  	\vspace{-1.5cm}
  \scriptsize
   \renewcommand\arraystretch{0.8}
	\setlength{\abovecaptionskip}{0pt}
	\setlength{\belowcaptionskip}{10pt}
	\begin{threeparttable}[b]
		\caption{Summary of Study 2, with smooth-varying coefficients within each cluster.} 
		\begin{tabular}{p{1.8cm}<{\centering}p{1.5cm}<{\centering}p{1.5cm}<{\centering}p{1.5cm}<{\centering}p{1.5cm}<{\centering}p{1.3cm}<{\centering}p{1.3cm}<{\centering}p{1cm}<{\centering}p{1cm}<{\centering}} 
			\hline
			\hline
			\multicolumn{1}{p{1.8cm}<{\centering}}{ Pattern}&\multicolumn{1}{p{1.5cm}<{\centering}}{Correlation} & \multicolumn{1}{p{1.5cm}<{\centering}}{Methods}& \multicolumn{1}{p{1.5cm}<{\centering}}{$\text{MSE}_{\beta_2}$}&\multicolumn{1}{p{1.5cm}<{\centering}}{$\text{MSE}_{\beta_1}$}&\multicolumn{1}{p{1.25cm}<{\centering}}{$\text{RI}_2$}&\multicolumn{1}{p{1.25cm}<{\centering}}{$\text{RI}_1$}&\multicolumn{1}{p{1.25cm}<{\centering}}{$\text{IC}_2$} &\multicolumn{1}{p{1.25cm}<{\centering}}{$\text{IC}_1$}\\
			\cmidrule(lr){4-5}\cmidrule(lr){6-7}\cmidrule(lr){8-9}
			&&SCVC&0.025&0.011&99.73&99.91&5.60&5.00\\
			&&&(0.001)&(0.001)&(0.02)&(0.01)&(0.06)&(0.00)\\
			&&SCC&0.279&0.348&68.22&68.26&151.08&120.60\\
			&&&(0.008)&(0.009)&(0.06)&(0.05)&(2.68)&(1.74)\\
			&weak&SCC*&0.254&0.390&70.26&75.28&53.34&35.15\\
			&&&(0.001)&(0.001)&(0.03)&(0.02)&(0.14)&(0.11)\\
			&&GWR&0.235&0.256&-&-&-&-\\
			&&&(0.005)&(0.005)&-&-&-&-\\
			&&PSE&0.828&0.854&-&-&-&-\\
			MST-equal&&&(0.002)&(0.002)&-&-&-&-\\
			\cmidrule(lr){3-9}
			&&SCVC&0.085&0.083&98.71&99.30&4.61&4.11\\
			&&&(0.009)&(0.011)&(0.27)&(0.21)&(0.07)&(0.03)\\
			&&SCC&1.663&1.611&69.36&68.91&131.49&101.80\\
			&&&(0.027)&(0.031)&(0.04)&(0.05)&(2.18)&(1.30)\\
			&strong&SCC*&1.456&1.300&70.52&75.34&49.92&45.50\\
			&&&(0.009)&(0.009)&(0.04)&(0.02)&(0.12)&(0.16)\\
			&&GWR&1.906&2.887&-&-&-&-\\
			&&&(0.024)&(0.049)&-&-&-&-\\
			&&PSE&1.681&1.437&-&-&-&-\\
			&&&(0.008)&(0.008)&-&-&-&-\\
			\hline
			&&SCVC&0.071&0.062&85.47&84.73&8.04&8.86\\
			&&&(0.001)&(0.002)&(0.02)&(0.01)&(0.02)&(0.07)\\
			&&SCC&0.349&0.579&71.90&72.43&172.80&124.43\\
			&&&(0.008)&(0.009)&(0.06)&(0.05)&(2.68)&(1.74)\\
			&weak&SCC*&0.352&0.528&73.83&75.90&57.21&63.36\\
			&&&(0.001)&(0.002)&(0.01)&(0.01)&(0.11)&(0.15)\\
			&&GWR&0.515&0.836&-&-&-&-\\
			&&&(0.003)&(0.005)&-&-&-&-\\
			&&PSE&1.107&1.447&-&-&-&-\\
			MST-unequal&&&(0.002)&(0.003)&-&-&-&-\\
			\cmidrule(lr){3-9}
			&&SCVC&0.580&1.390&81.17&81.97&6.79&7.38\\
			&&&(0.020)&(0.064)&(0.17)&(0.21)&(0.11)&(0.12)\\
			&&SCC&2.226&4.312&72.18&73.94&170.86&83.16\\
			&&&(0.025)&(0.049)&(0.02)&(0.08)&(1.82)&(0.94)\\
			&strong&SCC*&1.972&3.907&73.61&79.03&53.58&26.95\\
			&&&(0.005)&(0.013)&(0.01)&(0.01)&(0.14)&(0.04)\\
			&&GWR&4.989&4.387&-&-&-&-\\
			&&&(0.131)&(0.047)&-&-&-&-\\
			&&PSE&2.485&2.705&-&-&-&-\\
			&&&(0.008)&(0.018)&-&-&-&-\\
			\hline
			\hline
		\end{tabular}
		\label{table1}
		\begin{tablenotes} 
			\item SCVC: spatially clustered varying coefficient method; SCC: spatially clustered coefficient regression based on LASSO; SCC*: spatially clustered coefficient regression based on SCAD; GWR: geographically weighted regression; PSE: $P$-spline estimator. $\text{MSE}_{\beta_k}$/$\text{RI}_k$/$\text{IC}_k$: mean squared error $(\times 10)$/rand index $(\times 100)$/number of identified clusters, for $k$-th covariate, $k=1, 2$. Values in the parentheses are the standard errors. Note that GWR and PSE can not identify clusters.
		\end{tablenotes}
	\end{threeparttable}
\end{table}

\vspace{-0.7cm}
\section{Water Mass Analysis}\label{real-analysis}
\vspace{-0.4cm}
In oceanography,  
water masses detection is important, as it strongly affects the {ocean current} and global climate system \citep{nandi2004seismic, talley2011descriptive}.  The water masses are usually identified through the T-S relationship, because the T-S relationship is likely to change rapidly across the narrow boundaries (termed as $fronts$ in geoscience) between adjacent fluid masses \citep{li2019spatial}.  To study the T-S relationship and meanwhile detect different water masses, we apply the proposed SCVC method. For comparison, the SCC, SCC*, GWR, and PSE methods are also included. {The implementation of these five methods is the same as that in the simulation studies.}

\begin{figure}[htbp]
	\centering
	\subfigure{\hspace{-4.8cm}
	\begin{minipage}[t]{0.6\linewidth}		
	\includegraphics[width=3.7in,height=2.7in]{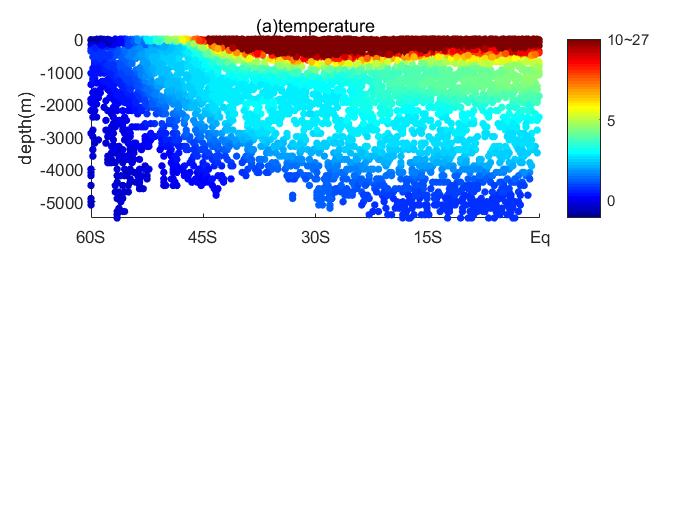}
    \end{minipage}%
     }%
 	\hspace{-1cm}
 	\subfigure{
 	\begin{minipage}[t]{0.25\linewidth}		
 		\includegraphics[width=3.7in,height=2.7in]{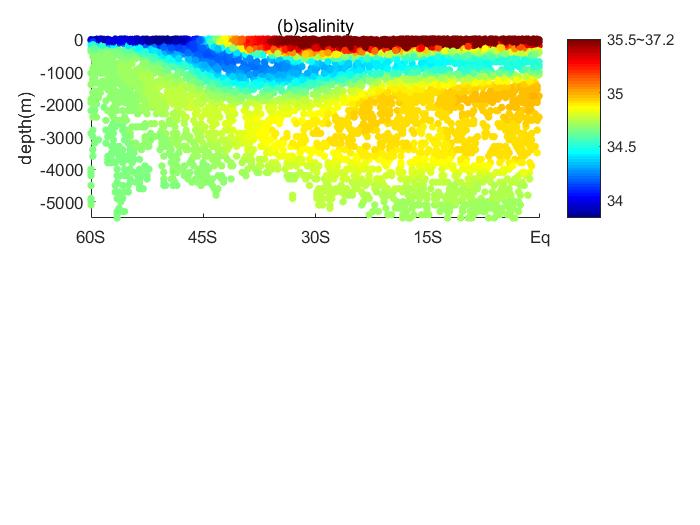}
 	\end{minipage}%
    }%
    \vspace{-4cm}
	\caption{  Spatial distributions of (a) temperature($^\circ$C) and (b) salinity(PSU) along $25^\circ$W.} 
	\label{Figure5}
\end{figure}

The data set contains 5130 observations of temperature and salinity in the Southern Hemisphere, along $25^\circ$W between $60^\circ$S and the equator ($0^\circ$),  see Figure \ref{Figure5}. This data set can be obtained from the World Atlas 2013, version 2 (WOA 13 V2),  archived at the National Oceanographic Data Center (\url{https://www.nodc.noaa.gov/OC5/woa13/}).   From Figure \ref{Figure5}, we find that the temperature is generally higher in the upper ocean and at lower latitudes, as a result of solar radiation, while the spatial structure of salinity is more complicated. 

To study the T-S relationship, we consider the following regression model,
\begin{equation*}
Sa(\bm{s}_i)=Te(\bm{s}_i)\beta(\bm{s}_i)+\beta_0(\bm{s}_i)+\epsilon({\bm{s}_i}),
\end{equation*}
where $Sa(\bm{s}_i)$ is the salinity at location $\bm{s}_i=(s_{ih}, s_{iv})$, $|s_{ih}|$ represents the horizontal distance (km) to the equator, and $|s_{iv}|$ is the vertical distance (km) to the sea surface; $Te(\bm{s}_i)$ is the temperature, $\beta(\bm{s}_i)$ measures the T-S relationship of interest, $\beta_0(\bm{s}_i)$ is the intercept. Notice the magnitude of depth and width of the ocean are quite different,  leading to  strong anisotropy. To alleviate this,  a common practice in oceanic studies \citep{vallis2017atmospheric} is to replace $\bm{s}_i$ by $(s_{ih}/H, s_{iv}/V)$,  where $H(V)$ is the horizontal (vertical) length of the ocean. 

\begin{figure}[htbp]
	\centering
	\vspace{-1cm}
	\subfigure{\hspace{-4.5cm}
		\begin{minipage}[t]{0.6\linewidth}
		
		\includegraphics[width=4in,height=3in]{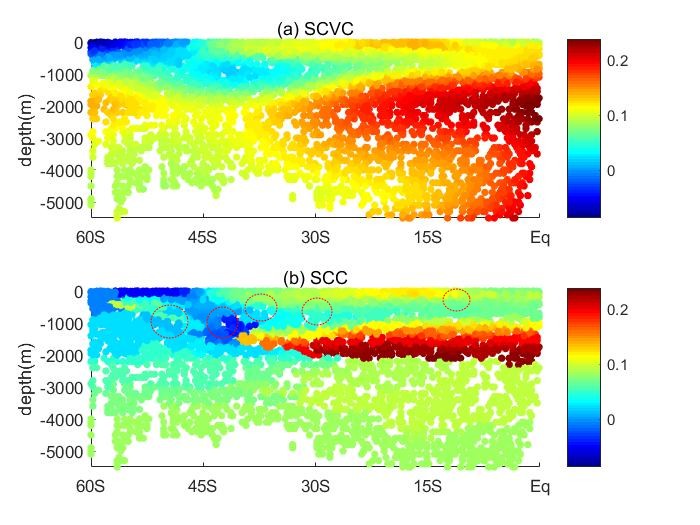}
		\includegraphics[width=4in,height=3in]{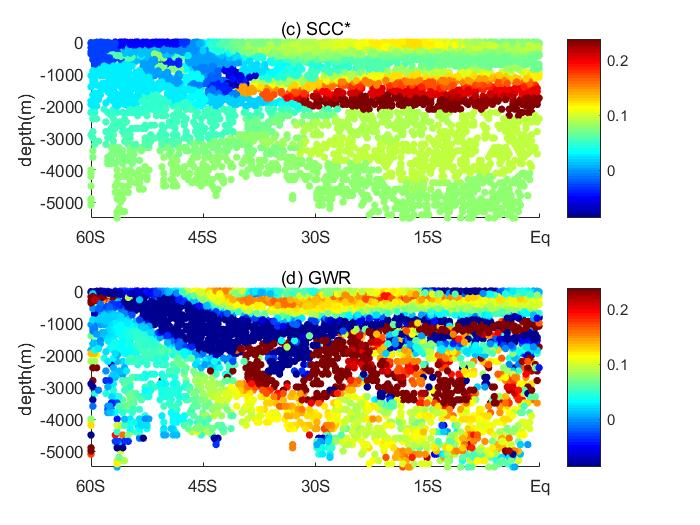}
	    \includegraphics[width=4in,height=3in]{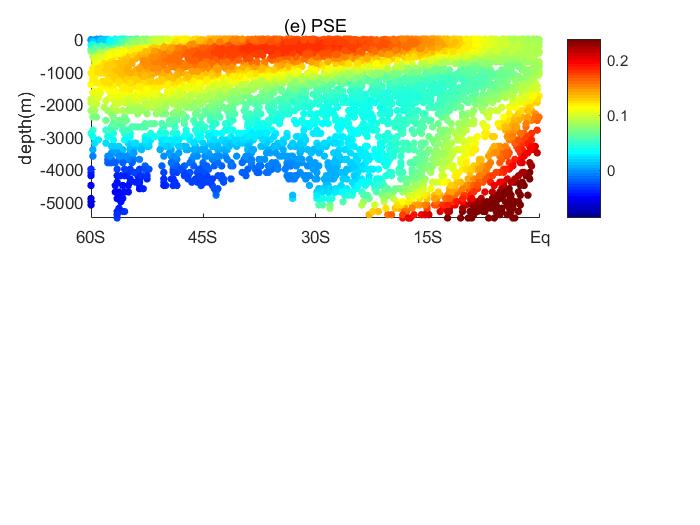}
       \vspace{-5cm} 
			\caption*{(i)}
		\end{minipage}%
	}%
	\subfigure{
		\begin{minipage}[t]{0.25\linewidth}
		\includegraphics[width=3.7in,height=3in]{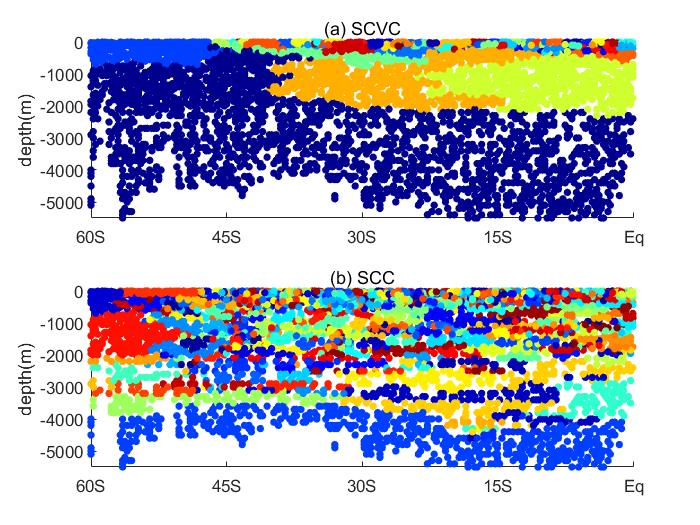}
		\vspace{-1cm}\includegraphics[width=3.7in,height=3in]{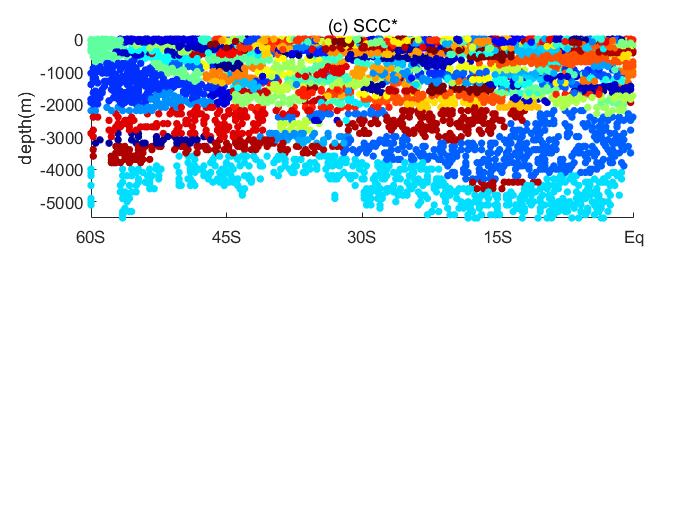}
		\vspace{-3.8cm}
		\caption*{\hspace{4.5cm}(ii)}
		\end{minipage}%
	}%
	\centering
	\caption{ Five figures in (i) represent the T-S relationship $\beta(\bm{s}_i)$ estimated by (a) SCVC,  (b) SCC, (c) SCC*,  (d) GWR and (e) PSE, { where the  colors in (i) represent the values of $\beta(\bm{s}_i)$}; {circles in (b) contain some small areas where the estimated coefficients from SCC vary.}  Three figures in (ii) represent the clustered pattern of estimated $\beta(\bm{s}_i)$ from  (a) SCVC,  (b) SCC and (c) SCC*, {where different colors in (ii) represent different clusters.}}
	\label{figure6}
\end{figure}

Figure \ref{figure6}(i) shows the estimated coefficient $\beta(\bm{s}_i)$  from SCVC, SCC, SCC*,  GWR and PSE. 
First, for the results from SCVC, the value of  $\beta(\bm{s}_i)$ is generally higher when the location $\bm{s}_i$ is closer to the equator, which is possibly due to the fact that the salinity in Figure \ref{Figure5} is generally higher for the locations closer to the equator.  Moreover, the value of  $\beta(\bm{s}_i)$ is negative when the location $\bm{s}_i$ is between $45^\circ$S and $60^\circ$S, with depth from $0$m to $-700$m. One possible explanation is, the salinity is quite low for this area, as observed in Figure \ref{Figure5}.  
Second, for the results from SCC and SCC*, the estimated coefficient $\beta(\bm{s}_i)$ is lower between $30^\circ$S and $0^\circ$ than that of SCVC, and the negative values of $\beta(\bm{s}_i)$  between $45^\circ$S and $60^\circ$S do not form a clear cluster. Moreover, it can be observed that, the estimated coefficients are more likely to vary even in a small area (there are many such areas, and we circle some of them in Figure \ref{figure6}(i)-(b)), indicating its assumption of constancy within each subregion is violated.
Third, for the results from GWR, the estimated coefficient  $\beta(\bm{s}_i)$ is quite noisy, even in the abyssal ocean, which is not consistent with the fluid dynamics, because over such a short distance in the abyssal ocean, there is no dynamical process that can lead to changes of the T-S relationship \citep{talley2011descriptive}. {Lastly, for the  results from PSE, the estimated coefficient  $\beta(\bm{s}_i)$ is generally positive/negative between $45^\circ$S and $60^\circ$S, with depth from $0$m to $-500$m/ $-4000$m to $-5000$m, which are opposite to the results of other methods, and inconsistent with the fact that the salinity is quite low for the former area (indicating negative T-S relationship) and relatively high for the latter area, see Figure \ref{Figure5}.  }

Figure \ref{figure6}(ii) shows a clearer comparison of SCVC,  SCC and SCC* {in clusters' detection}. 
The clustered patterns from SCC and SCC* are quite noisy, and fail to identify the water masses. 
Using the SCVC method instead, we obtain a much clearer clustered pattern shown in Figure \ref{figure6}(ii)-(a). 
First, from the bottom of the ocean to the surface, the number of identified clusters increases. This is consistent with the properties of salinity and temperature, whose variation is more severe near the sea surface \citep{emery2001water}. 
Second, the largest cluster between $60^\circ$ S and $0^\circ$, with depth from around $-2000$m to $-5500$m, corresponds to the North Atlantic Deep Water \citep{emery2001water},  which is essential to the Atlantic Meridional Overturning Circulation (AMOC) \citep{schmittner2007introduction}. 
Last, the cluster between $60^\circ$S and $45^\circ$S, with depth from $0$m to $-700$m,  corresponds to the Antarctic Surface Water \citep{florindo2008antarctic}, whose salinity is quite low, see Figure \ref{Figure5}.

\vspace{0.1cm}
\noindent {\large\bf Supplementary Materials}

Online Supplement contains the technical assumptions,  technical proofs of Theorems 1-3, the Nelder–Mead algorithm for minimizing $\text{BIC}(\{\lambda_k, \varrho_k\}_{k=1}^p)$,  how to generate the spatially clustered pattern in the simulation studies, {the results of SCC* with the initial values set as the true values,} {and additional simulation studies with smooth-varying coefficients}.

\noindent {\large\bf Acknowledgments}

This work is partially supported by National Natural Science Foundation of China grants 11671096, 11690013,  11731011 and  11871376.

\newpage
\begin{center}
	\Large{\bf Supplementary to ``Spatially Clustered Varying Coefficient Model''}
\end{center}

\setcounter{equation}{0}
\setcounter{figure}{0}
\setcounter{table}{0}
\setcounter{page}{1}
\setcounter{section}{0}
\makeatletter
\def\theequation{S.\arabic{equation}}
\def\thelemma{S.\arabic{lemma}}
\def\thesection{S\arabic{section}}	
\def\thefigure{S\arabic{figure}}
\def\thetable{S\arabic{table}}	

The online Supplementary Materials contain the technical assumptions,  technical proofs of Theorems 1-3,  the Nelder–Mead algorithm for minimizing $\text{BIC}(\{\lambda_k, \varrho_k\}_{k=1}^p)$,  how to generate the spatially clustered pattern in the simulation studies, {the results of SCC* with the initial values set as the true values,}  {and additional simulation studies with smooth-varying coefficients}.

\section{Technical assumptions}
\label{Sect1}
For any $s\times t$ matrix $\bm{A}=(A_{ij})_{i=1, j=1}^{s, t}$, denote $\|\bm{A}\|_1=\underset{1\leq j\leq t}{\text{max}} \sum_{i=1}^s |A_{ij}|$. Denote $\widetilde{\bm{Z}}^k=(\widetilde{\mathbb{Z}}_1^k, \cdots, \widetilde{\mathbb{Z}}_L^k)\in \mathbb{R}^{n\times Ln}$, $\widetilde{\mathbb{Z}}_l^k=\text{diag}(Z_{l, 1}^k, \cdots, Z_{l, n}^k)$ $\in \mathbb{R}^{n\times n}$, $l=1, \cdots, L$, $k=1, \cdots, p$,  where $Z_{l, i}^k=x_k(\bm{s}_i)B_l(\bm{s}_i) \in \mathbb{R}, i=1, \cdots, n$.   For $t\in \mathbb{R}$ and $\bm{t}=(t_1, \cdots, t_L)^T\in \mathbb{R}^L$, denote
\begin{equation*}
\rho_k(t)=\lambda_k^{-1} P_{\lambda_k}(t)\ \text{and} \ \bar{\rho}_{k, l}(\|\bm{t}\|_2)=\rho_k'(\|\bm{t}\|_2)\text{Sgn}({t_l}/{\|\bm{t}\|_2}),\ l=1, \cdots, L,
\end{equation*}
where $\text{Sgn}({t_l}/{\|\bm{t}\|_2})={t_l}/{\|\bm{t}\|_2}$ for $ \|\bm{t}\|_2\neq 0$; {for $\|\bm{t}\|_2=0$, $(\text{Sgn}({t_1}/{\|\bm{t}\|_2}),\ldots,\text{Sgn}({t_L}/{\|\bm{t}\|_2}))^T$ is any vector with $L_2$ norm less than 1. {For any square matrix $\bm{C}$,  $\lambda_\text{min}(\bm{C})$ and $\lambda_\text{max}(\bm{C})$ represent the smallest and largest eigenvalues of $\bm{C}$, respectively.}
	
	We recall the definition of $\mathcal{N}_{\mathcal{G}}^k$ in the beginning of Section 3.2 of main paper here, that is, 
	\begin{equation*}
	\mathcal{N}_{\mathcal{G}}^k=\{\bm{a}_k=(\bm{a}_{k, 1}^T, \cdots, \bm{a}_{k, n}^T)^T\in \mathbb{R}^{nL}: \bm{a}_{k, i}=\bm{a}_{k, j}~\text{for any}\  i, j\in \mathcal{M}^{g_k^*}_{k}, 1\leq g_k^*\leq G_k^* \}. 
	\end{equation*}
	For  $\bm{a}_k\in \mathcal{N}_{\mathcal{G}}^k$, let $\widetilde{\bm{\alpha}}_{k, g_k^*}=(\widetilde{\alpha}_{ k, g_k^*}^1, \cdots, \widetilde{\alpha}_{ k, g_k^*}^L)^T\in\mathbb{R}^L$, {where $\widetilde{\bm{\alpha}}_{k, g_k^*}=\bm{a}_{k, i},\  \forall i\in \mathcal{M}^{g_k^*}_{k}$.} Denote ${\bm{\alpha}}_k=({\bm{\alpha}}_{k, 1}^T, \cdots, {\bm{\alpha}}_{k, L}^T)^T\in \mathbb{R}^{LG_k^*}$, $k=1, \cdots, p$,  where ${\bm{\alpha}}_{k, l}=(\widetilde{\alpha}_{ k,  1}^l, \cdots, \widetilde{\alpha}_{k,   G_k^* }^l)^T$ $ \in \mathbb{R}^{G_k^*}$, $l=1, \cdots, L$.  According to the definition of $\hat{\bm{a}}^{\text{or}}=((\hat{\bm{a}}_1^{\text{or}})^T, \cdots, (\hat{\bm{a}}_p^{\text{or}})^T)^T$ in Section 3.2 of the main paper,  the oracle estimator for $\bm{\alpha}=(\bm{\alpha}_1^T, \cdots, \bm{\alpha}_p^T)^T\in \mathbb{R}^{L(\sum_{k=1}^p G_k^*)}$  is
	\begin{equation}
	\hat{\bm{\alpha}}^{or}=\underset{\bm{\alpha}\in \mathbb{R}^{L(\sum_{k=1}^p G_k^*)} }{\text{arg}\  \text{min}} \frac{1}{2n} \|\bm{Y}- \bm{Z}  \bm{\alpha}\|_2^2+ \bm{\alpha}^T \bm{\Pi}\bm{\alpha},
	\label{Eq1}
	\end{equation}
	where $\bm{Z}=(\bm{Z}^1, \cdots, \bm{Z}^p)$,  {$\bm{Z}^k=(\mathbb{Z}_1^k, \cdots, \mathbb{Z}_L^k)\in \mathbb{R}^{n\times LG_k^*}$} and $\mathbb{Z}_l^k=\widetilde{\mathbb{Z}}_l^k\bm{\Lambda}_k\in \mathbb{R}^{n\times G_k^*}$, $l=1, \cdots, L$, with   $\bm{\Lambda}_k=\{\Lambda_{i\,{g_k^*}}\}$ being the ${n\times G_k^*}$ matrix with $\Lambda_{i\, {g_k^*}}=1$ for $i\in \mathcal{M}^{g_k^*}_{k}$ and $\Lambda_{i\, {g_k^*}}=0$ otherwise;
	$\bm{\Pi}=\text{diag}(\varrho_1\bm{\Pi}_1, \cdots, \varrho_p \bm{\Pi}_p)$, where $\bm{\Pi}_k=\text{diag}(h_{11}\widetilde{\bm{\Lambda}}_k, \cdots, h_{LL}\widetilde{\bm{\Lambda}}_k)$ $\in \mathbb{R}^{LG_k^*}$, $\widetilde{\bm{\Lambda}}_k=\bm{\Lambda}_k^T\bm{\Lambda}_k=\text{diag}(|\mathcal{M}^{1}_{k}|, \cdots, |\mathcal{M}^{G_k^*}_{k}|)$, with $|\mathcal{M}^{g_k^*}_{k}|$ be the number of elements in $\mathcal{M}^{g_k^*}_{k}$, $k=1, \cdots, p$,  and  $h_{ll}$ is the $l-$th diagonal element of $\PRC$ with $h_{ll}=0$, $l=1, 2, 3$ and $h_{ll}=1$, $l=4, \cdots, L$.
	
	We make the following assumptions.
	\begin{enumerate}[label=(A\arabic*)]

		\item  $\lambda_{\text{min}}(\bm{Z}^T\bm{Z})\geq C_1 |\mathcal{M}_\text{min}|$ for a positive constant $C_1$.
		\label{Au1}
		\item Let $\mathbb{Z}_{l, g_k^*}^k$ be the $g_k^*-$th column of $\mathbb{Z}_l^k$, and  assume {$\|\mathbb{Z}_{l, g_k^*}^k\|_2\leq C_2 \sqrt{ |\mathcal{M}^{g_k^*}_{k}|}$, $k=1, \cdots, p$, $g_k^*=1, \cdots, G_k^*$,} for some positive constant  $C_2\geq 1$.
		\label{Au2}
		\item $\|\widetilde{\mathbb{Z}}_l^k\|_\infty< C_3$, $k=1, \cdots, p$, $l=1, \cdots, L$, for some positive constant $C_3$.
		\label{Au5}
		\item {For any $k\in \{1, \cdots, p\}$, $P_{\lambda_k}(t)$, i.e., the penalty function in (4) of the main paper},  is a symmetric function of $t$, which is nondecreasing and concave in $t\in [0,\ \infty)$. There exists a positive constant $a_0$ such that $\rho_k(t)$ is constant for all $t\geq a_0\lambda_k$, and $\rho_k(0)=0$, $\rho_k'(t)$ exists and is continuous except for a finite number of $t$ and $\rho_k'(0_{+})=1$.
		\label{Au4}
		\item The noise vector $\bm{\epsilon}=\{\epsilon(\bm{s}_1), \cdots, \epsilon(\bm{s}_n)\}^T$ has sub-Gaussian tails such that $P(|\bm{a}^T\bm{\epsilon}|>\|\bm{a}\|_2x)\leq 2 \text{exp}(-c_1x^2)$ for any  vector $\bm{a}\in \mathbb{R}^n$ and $x>0$, where $0<c_1<\infty$.
		\label{Au6}
		\item Each subregion $\mathcal{D}_{k}^{g_k}$ contains a open set,    $g_k=1, \cdots, G_k$, $k=1, \cdots, p$.
		\label{Au7}
	\end{enumerate} 
	Assumptions \ref{Au1}-\ref{Au5} are regular conditions in the context of subgroup analysis. Assumption \ref{Au1} is similar with Assumption (C1) in \cite{ma2017concave}.  Assumptions \ref{Au2}-\ref{Au5} can be easily satisfied under infill domain. It is because,  under infill domain,  all the locations $\bm{s}_i$, $i=1, \cdots, n$,  are within a bounded domain, indicating that $B_l(\bm{s}_i)$,  $l=1, \cdots, L$ are bounded. Thus, Assumptions \ref{Au2}-\ref{Au5} are satisfied,  when the covariate $x_k(\bm{s}_i)$, $i=1, \cdots, n$ are bounded.  Assumption \ref{Au4} are satisfied for common concave penalties such as MCP and SCAD. Assumption \ref{Au6} is commonly assumed in high-dimensional settings. Assumption \ref{Au7} is quite weak, indicating that the area of each subregion is not zero.
	
	\section{Technical proofs}
	\begin{lemma}\label{lemma1}
		For the basis functions $\bm{B}(\bm{s})$ defined in Section 2.3.2, we have, $\bm{d}_1^T\bm{B}(\bm{s})\neq \bm{d}_2^T\bm{B}(\bm{s})$ for some $\bm{s}\in\mathcal{D}_{k}^{g_k}$, if and only if $\bm{d}_1\neq \bm{d}_2$.
	\end{lemma}
	{\bf{Proof of Lemma \ref{lemma1}}}.
	First, we prove that, if $\bm{d}_1^T \bm{B}(\bm{s})\neq \bm{d}_2^T\bm{B}(\bm{s})$ for some $\bm{s}\in  \mathcal{D}_{k}^{g_k}$,  we have $\bm{d}_1\neq \bm{d}_2$. This is obvious.
	
	Second, we prove that, if $\bm{d}_1\neq \bm{d}_2$,  we have $\bm{d}_1^T \bm{B}(\bm{s})\neq \bm{d}_2^T\bm{B}(\bm{s})$ for some $\bm{s}$ over $\mathcal{D}_{k}^{g_k}$.  It is equivalent to prove that, if 	$\bm{d}_1^T \bm{B}(\bm{s})= \bm{d}_2^T\bm{B}(\bm{s})$ for all $\bm{s}$ over $\mathcal{D}_{k}^{g_k}$, we have $\bm{d}_1= \bm{d}_2$. 
	Noticing the basis functions $\bm{B}(\bm{s})$, defined in Section 2.3.2,  are analytic over $\mathcal{D}$, according to the uniqueness of analytic continuation and Assumption \ref{Au7}, we know that,  $\bm{d}_1^T \bm{B}(\bm{s})= \bm{d}_2^T\bm{B}(\bm{s})$ for all $\bm{s}$ over $\mathcal{D}_{k}^{g_k}$, if and only if,  $\bm{d}_1^T \bm{B}(\bm{s})= \bm{d}_2^T\bm{B}(\bm{s})$ for all $\bm{s}$ over $\mathcal{D}$. Thus, we only need to prove that, if $\bm{d}_1^T \bm{B}(\bm{s})= \bm{d}_2^T\bm{B}(\bm{s})$ for all $\bm{s}$ over $\mathcal{D}$, we have $\bm{d}_1= \bm{d}_2$. 
	By $\bm{d}_1^T \bm{B}(\bm{s})= \bm{d}_2^T\bm{B}(\bm{s})$ over $\mathcal{D}$, we have 
	\begin{equation}
	\bm{d}_1^T \int_{\mathcal{D}}\bm{B}(\bm{s}) \bm{B}(\bm{s})^T d\bm{s}= \bm{d}_2^T\int_{\mathcal{D}}\bm{B}(\bm{s}) \bm{B}(\bm{s})^T d\bm{s}.
	\label{eqnnnn1}
	\end{equation}
	Because for valid basis functions, it is the basic requirement that $\int_{\mathcal{D}}\bm{B}(\bm{s}) \bm{B}(\bm{s})^T d\bm{s}$ is positive definite, see \cite{zhou1998local}. Thus, by multiplying its inverse matrix in both sides of (\ref{eqnnnn1}),    we have $\bm{d}_1=\bm{d}_2$. Proof is completed.   $\quad\square$
	\vspace{0.5cm}
	
	{\bf{Proof of Theorem 1}}.
	To prove that the set of true clusters $\{\mathcal{M}^{1}_{k}, \cdots, \mathcal{M}^{G_k^*}_{k}\}$ exists and is unique, we only need to prove that for the  $\mathcal{G}^{g_k}_k$, in which 
	not all the location pairs  can be connected through a
	path, made up of the edges in $\mathbb{E}_{k}^{g_k}$,
	the partition  $\mathcal{G}^{g_k}_{k, 1},\cdots, \mathcal{G}^{g_k}_{k, F_k}$ exists and is unique.
	
	For the proof of existence, we randomly select a starting location $\bm{s}_i$, $i\in \mathcal{G}^{g_k}_k$, then  through the edges in $\mathbb{E}_{k}^{g_k}$, we can find the location index set $\mathcal{L}_1$ containing $i$, satisfying that, for any two locations $\bm{s}_{i_1}$ and $\bm{s}_{i_2}$,  $i_1\in \mathcal{L}_1$, $i_2\in \mathcal{L}_1$, they are connected by a path, made up of  edges in $\mathbb{E}_{\mathcal{L}_1}=\{(i, j):(i, j)\in \mathbb{E}_{\text{MST}},\ \text{and}\  i, j\in \mathcal{L}_1\}$, and for any two locations $\bm{s}_{i_3}$, ${i_3}\in \mathcal{L}_1$, and $\bm{s}_{i_4}$, $i_4\in \mathcal{G}^{g_k}_k \setminus  \mathcal{L}_1 $, they satisfy the edge $(i_3, i_4)\notin \mathbb{E}_{\text{MST}}$. Similarly, we can construct $\mathcal{L}_2$ by randomly selecting a starting location $\bm{s}_{i_5}$, $i_5\in \mathcal{G}^{g_k}_k\setminus  \mathcal{L}_1.$ Repeat the constructing process until  $\mathcal{G}^{g_k}_k=\mathcal{L}_1\cup \cdots \cup \mathcal{L}_{F_k}$, and let $\mathcal{G}^{g_k}_{k, f}=\mathcal{L}_{f}, f=1, \cdots, F_k$. By the constructing process, we have that, for any two locations in $\mathcal{G}^{g_k}_{k, f}$, $f=1\cdots, F_k$, they are connected by a path, made up of some edges in $\mathbb{E}_{k, f}^{g_k}=\{(i, j):(i, j)\in \mathbb{E}_{\text{MST}},\ \text{and}\  i, j\in \mathcal{G}^{g_k}_{k, f}\}$; { meanwhile, for any two locations $\bm{s}_{i}\in \mathcal{G}^{g_k}_{k, f}$, $\bm{s}_{j}\in \mathcal{G}^{g_k}_{k, f'}$ and $f\neq f'$, {the corresponding edge $(i, j)\notin \mathbb{E}_{\text{MST}}$;}}  Thus, the existence is  proved.
	
	For the proof of uniqueness,  it is equivalent to prove that for any starting locations $\bm{s}_i, i\in \mathcal{G}^{g_k}_k$, the resulting partition is still $\mathcal{L}_1, \cdots, \mathcal{L}_{F_k}$ with repeating the above process. Thus, we only need to prove that, for the starting point $\bm{s}_i$, $i\in \mathcal{L}_f$, $f\in\{1, \cdots, F_k\}$, by the above process, the first partition is $\mathcal{L}_f$. The proof is quite  straightforward by the property of $\mathcal{L}_f$, thus is omitted here.
	
	From the above discussion,  proof is completed. Moreover, the  proof for the existence and uniqueness of the set of estimated clusters $\{\widehat{\mathcal{M}}^{1}_{k}, \cdots, \widehat{\mathcal{M}}^{\widehat{G_k^*}}_{k}\}$ is similar, thus is omitted.  $\quad\square$
	\vspace{0.5cm}

	{\bf{Proof of Theorem 2}}: According to (\ref{Eq1}), we have
	\begin{equation*}
	\hat{\bm{\alpha}}^{or}
	=(\bm{Z}^T\bm{Z}+2n{\bm{\Pi}})^{-1}\bm{Z}^T \bm{Y}.
	\end{equation*}
	Let $\bm{\Omega}=\bm{Z}^T\bm{Z}$, and $\bm{\alpha}^0$ represent the true value of $\bm{\alpha}$,   then 
	\begin{equation}
	\begin{split}
	\hat{\bm{\alpha}}^{or}-{\bm{\alpha}}^0
	=\left\{\bm{\Omega}+2n{\bm{\Pi}}\right\}^{-1}\bm{Z}^T \bm{\epsilon}
	+\left\{\left(\bm{\Omega}+2n{\bm{\Pi}}\right)^{-1}-\bm{\Omega}^{-1}\right\}\bm{\Omega}{\bm{\alpha}}^0:=\bm{I}_1+\bm{I}_2.
	\label{eqnnn1}
	\end{split}
	\end{equation}
	For $\bm{I}_1$, we have 
	\begin{equation}
	\|\bm{I}_1\|_\infty\leq \|\left\{\bm{\Omega}+2n{\bm{\Pi}}\right\}^{-1}\|_\infty \|\bm{Z}^T \bm{\epsilon}\|_\infty.
	\label{eqnn1}
	\end{equation}
	By Assumption \ref{Au1}, we have $\|(\bm{\Omega}+n\lambda_2{\bm{\Pi}})^{-1}\|_\infty\leq \sqrt{L(\sum_{k=1}^p G_k^*)}C_1^{-1}|\mathcal{M}_\text{min}|^{-1}$. Moreover,  for some constant $c>0$,
	\begin{equation*}
	\begin{split}
	P(\|\bm{Z}^T \bm{\epsilon}\|_\infty>c\sqrt{n \text{log}\, n})\leq& \sum_{k=1}^p P(\|(\bm{Z}^k)^T \bm{\epsilon}\|_\infty>c\sqrt{n \text{log}\, n}),
	\end{split}
	\end{equation*}
	and by Assumptions \ref{Au2} and {\ref{Au6}},
	\begin{eqnarray} 
	&&P(\|(\bm{Z}^k)^T \bm{\epsilon}\|_\infty>c\sqrt{n \text{log}\, n})\leq \sum_{l=1}^L P(\|(\mathbb{Z}_l^k)^T \bm{\epsilon}\|_\infty>c\sqrt{n \text{log}\, n})\nonumber\\
	&\leq& \sum_{l=1}^L \sum_{g_k^*=1}^{G_k^*} P(|\sum_{i\in \mathcal{M}^{g_k^*}_{k}} Z_{l, i}^k\epsilon_i|>c\sqrt{n \text{log}\, n})\nonumber\\
	&\leq& \sum_l^L \sum_{g_k^*=1}^{G_k^*} P(|\sum_{i\in \mathcal{M}^{g_k^*}_{k}} Z_{l, i}^k\epsilon_i|>cC_2^{-1}\sqrt{\sum_{i\in \mathcal{M}^{g_k^*}_{k}} (Z_{l, i}^k)^2}\sqrt{\text{log}\, n})\nonumber\\
	&\leq& 2L\ (\sum_{k=1}^p G_k^*)\ \text{exp}(-c^2 c_1 C_2^{-2}\text{log}\, n).
	\label{sseqn1}
	\end{eqnarray}
	By (\ref{sseqn1}), taking $c=c_1^{-1/2}C_2$,   we have
	\begin{equation*}
	P(\|\bm{Z}^T \bm{\epsilon}\|_\infty>c\sqrt{n \text{log}\, n})\leq \frac{2pL (\sum_{k=1}^p G_k^*)}{n}.
	\end{equation*}
	Therefore, by (\ref{eqnn1}), with probability at least  $1- {2pL (\sum_{k=1}^p G_k^*)}/{n}$, 
	\begin{eqnarray}
	\|\bm{I}_1\|_\infty\leq \sqrt{L(\sum_{k=1}^p G_k^*)}C_1^{-1}|\mathcal{M}_\text{min}|^{-1}c_1^{-1/2}C_2\sqrt{n \text{log}\, n}.\label{I1-bound}
	\end{eqnarray}
	
	For $\bm{I}_2$, according to $\bm{A}^{-1}-\bm{B}^{-1}=\bm{B}^{-1}(\bm{B}-\bm{A})\bm{A}^{-1}$ for any invertible matrices $\bm{A}$ and $\bm{B}$, we have
	\begin{equation}
	\begin{split}
	&\|\bm{I}_2\|_\infty=\|\left\{\bm{\Omega}^{-1}-\left(\bm{\Omega}+2n{\bm{\Pi}}\right)^{-1}\right\}\bm{\Omega}{\bm{\alpha}}^0\|_\infty=\|\left(\bm{\Omega}+2n{\bm{\Pi}}\right)^{-1}(2n{\bm{\Pi}}){\bm{\alpha}}^0\|_\infty\\
	\leq& \|\left(\bm{\Omega}+2n{\bm{\Pi}}\right)^{-1}\|_\infty \|2n{\bm{\Pi}}\|_\infty \|{\bm{\alpha}}^0\|_\infty\leq \sqrt{L(\sum_{k=1}^p G_k^*)}C_1^{-1}|\mathcal{M}_\text{min}|^{-1}2n\widetilde{\varrho} |\mathcal{M}_\text{max}|\, \|{\bm{\alpha}}^0\|_\infty,\label{I2-bound}
	\end{split}
	\end{equation}
	where the second inequality comes from the definition of  ${\bm{\Pi}}$.
	
	Combing (\ref{eqnnn1}), (\ref{I1-bound}) and (\ref{I2-bound}), with probability at least  $1- {2pL (\sum_{k=1}^p G_k^*)}/{n}$, we have 
	\begin{equation*}
	\|\hat{\bm{\alpha}}^{or}-{\bm{\alpha}}^0\|_{\infty}\leq r_n,
	\end{equation*} 
	where $r_n=\sqrt{L(\sum_{k=1}^p G_k^*)}C_1^{-1}|\mathcal{M}_\text{min}|^{-1}\{c_1^{-1/2}C_2\sqrt{n \text{log}\, n}+2n\widetilde{\varrho} |\mathcal{M}_\text{max}|\, \|{\bm{\alpha}}^0\|_\infty\}$. Proof is completed. $\quad\square$.
	\vspace{0.5cm}
	
	{\bf{Proof of Theorem 3}}: 
	{We first introduce some notations which are frequently used in this proof.}
	For $1\leq g_k^*<{g_k^*}'\leq G_k^*$, let $\mathbb{C}_{ g_k^*{g_k^*}'}=\{(i,j)\in \mathbb{E}_{\text{MST}}: i \in \mathcal{M}^{g_k^*}_{k}\ \text{and}\ j \in \mathcal{M}^{{g_k^*}'}_{k}, \ \text{or}\  i \in \mathcal{M}^{{g_k^*}'}_{k}\ \text{and}\ j \in \mathcal{M}^{g_k^*}_{k}\} $, and $|\mathbb{C}_{ g_k^*{g_k^*}'}|$ denote the number of elements in $\mathbb{C}_{ g_k^*{g_k^*}'}$.
	Let $\mathcal{F}^k_{\mathcal{G}}$ be the subspace of $\mathbb{R}^{n}$, defined as
	\begin{equation*}
	\mathcal{F}_{\mathcal{G}}^k=\{\bm{b}=(b_1, \cdots, b_n)^T\in \mathbb{R}^{n}: b_i=  b_j,\ \text{for}\  \text{any}\  i, j\in \mathcal{M}^{{g_k^*}}_{k}, 1\leq g_k^*\leq G_k^* \}. 
	\end{equation*}
	Introducing the mapping $T_k: \mathcal{F}_{\mathcal{G}}^k\to \mathbb{R}^{G_k^*}$, where $T_k(\bm{b})$ is the $G_k^*-$dimensional vector, and its $g_k^*-$th coordinate equals to the common value of $b_i$ for $i\in \mathcal{M}^{{g_k^*}}_{k}$, {denoted as $b^{{g_k^*}}_k$}. Note that $T_k$ is a bijection and $T_k^{-1}$ is well-defined. Let $T^*_k: \mathbb{R}^n\to  \mathbb{R}^{G_k^*}$ be the mapping such that $T^*_k(\bm{b})=\{|\mathcal{M}^{{g_k^*}}_{k}|^{-1}\sum_{i\in \mathcal{M}^{{g_k^*}}_{k}} b_i\}_{g_k^*=1}^{G_k^*}$. It is easy to see that, for $\bm{b}\in \mathcal{F}_{\mathcal{G}}^k$, {we have $T^*_k(\bm{b})=T_k(\bm{b})=(b^{1}_k, \cdots, b^{G_k^*}_k)^T$.} {These mappings act as a bridge in the proving process to link $\bm{a}$ and $\bm{\alpha}$, where $\bm{a}$ is the unknown parameter vector in the objective function  (4) of the SCVC in the main paper, and $\bm{\alpha}$ is the unknown vector in the oracle procedure (\ref{Eq1}).} 
	
	{
		In the following part, there are many new notations based on $\bm{a}$ and $\bm{\alpha}$, we summarize them in this paragraph. Let ${a}_{k, i}^l$ be the $l-$th element of $\bm{a}_{k, i}$, $l=1, \cdots, L$, where $\bm{a}_{k, i}$, $k=1, \cdots, p$, $i=1, \cdots, n$,  is the spline coefficient vector in (4) of the main paper.    Consider the following matrix,
		\begin{equation}
		\begin{split}
		&a_{k, 1}^1,\ a_{k, 1}^2,\ \cdots,\ a_{k, 1}^L,\\
		&a_{k, 2}^1,\ a_{k, 2}^2,\ \cdots,\ a_{k, 2}^L,\\
		&\quad\quad\quad\quad\vdots\\
		&a_{k, n}^1,\ a_{k, n}^2,\ \cdots,\ a_{k, n}^L.
		\label{matrix}
		\end{split}
		\end{equation}
		Then, $\bm{a}_{k, i}$ is the $i-$th row of matrix (\ref{matrix}). We define the $l-$th column of  matrix (\ref{matrix}) as $\widetilde{\bm{a}}_{k, l}$, i.e., $\widetilde{\bm{a}}_{k, l}=({a}_{k, 1}^l, \cdots, {a}_{k, n}^l)^T$, $l=1, \cdots, L$. By the definition below (6) in the main paper, we have $\bm{a}=(\bm{a}_1^T, \cdots, \bm{a}_p^T)^T$, $\bm{a}_k=(\bm{a}_{k, 1}^T, \cdots, \bm{a}_{k, n}^T)^T$, $k=1, \cdots, p$. Similarly, we define $\widetilde{\bm{a}}=(\widetilde{\bm{a}}_1^T, \cdots, \widetilde{\bm{a}}_p^T)^T$, where $\widetilde{\bm{a}}_k=(\widetilde{\bm{a}}_{k, 1}^T, \cdots, \widetilde{\bm{a}}_{k, L}^T)^T$, $k=1, \cdots, p$. Thus, $\widetilde{\bm{a}}$ is a permutation of elements in $\bm{a}$. 
		Furthermore, we let $\widetilde{\bm{a}}_{k, l}^*=T^{-1}_k(T^*_k(\widetilde{\bm{a}}_{k, l}))$, which actually replaces each element in $\widetilde{\bm{a}}_{k, l}$ with the average over the corresponding SpaNeigh true cluster. We then denote $\widetilde{\bm{a}}_k^*=\{(\widetilde{\bm{a}}_{k, 1}^*)^T, \cdots, (\widetilde{\bm{a}}_{k, L}^*)^T\}^T$, $k=1, \cdots, p$, and
		$\widetilde{\bm{a}}^*=((\widetilde{\bm{a}}_1^*)^T, \cdots, (\widetilde{\bm{a}}_p^*)^T)^T$. Same as the relationship between ${\bm{a}}$ and $\widetilde{\bm{a}}$, we can define ${\bm{a}}^*$ based on $\widetilde{\bm{a}}^*$. Same as the relationship between ${\bm{a}}_k$/${\bm{a}}_{k, i}$ and ${\bm{a}}$, we can define $\bm{a}_k^*$ and $\bm{a}_{k, i}^*$ based on ${\bm{a}}^*$, i.e.,
		${\bm{a}}^*=(({\bm{a}}_1^*)^T, \cdots, ({\bm{a}}_p^*)^T)^T$ and $\bm{a}_k^*=({\bm{a}_{k, 1}^*}^T, \cdots, {\bm{a}_{k, n}^*}^T)^T$, $k=1, \cdots, p$.
		Now we recall the definition of $\bm{\alpha}$ above (\ref{Eq1}), the definition of $\bm{\alpha}$ is essentially based on $\bm{a}$ with knowing the information of the SpaNeigh true clusters. 
		For  $\bm{a}_k\in \mathcal{N}_{\mathcal{G}}^k$, 
		let $\widetilde{\bm{\alpha}}_{k, g_k^*}=(\widetilde{\alpha}_{ k, g_k^*}^1, \cdots, \widetilde{\alpha}_{ k, g_k^*}^L)^T\in\mathbb{R}^L$, {where $\widetilde{\bm{\alpha}}_{k, g_k^*}=\bm{a}_{k, i},\  \forall i\in \mathcal{M}^{g_k^*}_{k}$.}
		Then, we denote  ${\bm{\alpha}}_{k, l}=(\widetilde{\alpha}_{ k,  1}^l, \cdots, \widetilde{\alpha}_{k,   G_k^* }^l)^T$, $l=1, \cdots, L$, and let ${\bm{\alpha}}_k=({\bm{\alpha}}_{k, 1}^T, \cdots, {\bm{\alpha}}_{k, L}^T)^T$, $k=1, \cdots, p$, and $\bm{\alpha}=(\bm{\alpha}_1^T, \cdots, \bm{\alpha}_p^T)^T$. Moreover, whenever adding zero in the superscript of a symbol, it represents the corresponding true value, for example, $\bm{\alpha}^0$ is the true value of $\bm{\alpha}$.}

	{Based on the above notations,  we can write the objective function (4) of the SCVC with  $\mathbb{E}_{\text{MST}}$ in the main paper, in a compact form.} Let
	\begin{equation*}
	\begin{split}
	L_n(\widetilde{\bm{a}})&= \frac{1}{2n} \|\bm{Y}- \widetilde{\bm{Z}}  \widetilde{\bm{a}}\|_2^2+ \widetilde{\bm{a}}^T \bm{\Pi}^*\widetilde{\bm{a}},\ P_n(\widetilde{\bm{a}})={\sum_{k=1}^p\lambda_k\sum_{(i,j)\in \mathbb{E}_\text{MST}}\rho_k\left(\|\bm{a}_{k, i}-\bm{a}_{k, j}\|_2\right)},\\
	L_n^{\mathcal{G}}({\bm{\alpha}})&= \frac{1}{2n} \|\bm{Y}- \bm{Z}  \bm{\alpha}\|_2^2+ \bm{\alpha}^T \bm{\Pi}\bm{\alpha},\ P_n^{\mathcal{G}}({\bm{\alpha}})=\sum_{k=1}^p \lambda_k
	\underset{
		\begin{scriptsize}
		\begin{split}
		\vspace{-1cm}	 &\quad |\mathbb{C}_{ g_k^*{g_k^*}'}|\neq 0\\
		&1\leq g_k^*<{g_k^*}'\leq G_k^*
		\end{split}
		\end{scriptsize}
	}{\sum}
	{ |\mathbb{C}_{ g_k^*{g_k^*}'}| \rho_k\left(\|\widetilde{\bm{\alpha}}_{k, g_k^*}-\widetilde{\bm{\alpha}}_{k, {g_k^*}'}\|_2\right)},
	\end{split}
	\end{equation*}
	where  $\bm{\Pi}^*=\text{diag}(\varrho_1\bm{\Pi}^*_1, \cdots, \varrho_p\bm{\Pi}^*_p)$,  $\bm{\Pi}^*_k=\text{diag}(\underbrace{h_{11},\cdots, h_{11}}_{n}, \cdots, \underbrace{h_{LL},\cdots, h_{LL}}_{n})\in \mathbb{R}^{nL}$, $k=1, \cdots, p$, and $\widetilde{\bm{Z}}=(\widetilde{\bm{Z}}^1, \cdots, \widetilde{\bm{Z}}^p)$.
	Define
	\begin{equation*}
	Q_n(\widetilde{\bm{a}})=L_n(\widetilde{\bm{a}})+ P_n(\widetilde{\bm{a}}) ~\text{and}~  Q_n^{\mathcal{G}}({\bm{\alpha}})=L_n^{\mathcal{G}}({\bm{\alpha}})+P_n^{\mathcal{G}}({\bm{\alpha}}).
	\end{equation*}
	{Then, $Q_n(\widetilde{\bm{a}})$ is the compact form of the objective function (4) of the SCVC with  $\mathbb{E}_{\text{MST}}$ in the main paper, and $Q_n^{\mathcal{G}}({\bm{\alpha}})$ is defined based on $Q_n(\widetilde{\bm{a}})$ for technical purpose. }
	
	{We illustrate the relationship between $Q_n(\widetilde{\bm{a}})$ and $Q_n^{\mathcal{G}}({\bm{\alpha}})$.}  For every $\widetilde{\bm{a}}_{k, l}\in \mathcal{F}_{\mathcal{G}}^k$, $l=1, \cdots, L$, denote $\bm{T}_k(\widetilde{\bm{a}}_{k})=(T_k(\widetilde{\bm{a}}_{k, 1})^T, \cdots, T_k(\widetilde{\bm{a}}_{k, L})^T)^T$,  and $\bm{T}(\widetilde{\bm{a}})=(\bm{T}_1(\widetilde{\bm{a}}_{1})^T, \cdots, \bm{T}_p(\widetilde{\bm{a}}_{p})^T)^T$. By routine  calculation, we have $P_n^{\mathcal{G}}(\bm{T}(\widetilde{\bm{a}}))=P_n(\widetilde{\bm{a}})$. Moreover, for every $\bm{\alpha}_{k, l}\in \mathbb{R}^{G^*_k}$, $k=1, \cdots, p$, denote \\
	$\bm{T}^{-1}_k(\bm{\alpha}_k)=(T^{-1}_k({\bm{\alpha}}_{k, 1})^T, \cdots, T^{-1}_k({\bm{\alpha}}_{k, L})^T)^T$ and $\bm{T}^{-1}(\bm{\alpha})=(\bm{T}^{-1}_1(\bm{\alpha}_1)^T, \cdots, \bm{T}^{-1}_p(\bm{\alpha}_p)^T)^T$, then  we can obtain $P_n(\bm{T}^{-1}(\bm{\alpha}))=P_n^{\mathcal{G}}({\bm{\alpha}})$. Hence,
	\begin{equation}
	Q_n(\widetilde{\bm{a}})=Q_n^{\mathcal{G}}(\bm{T}(\widetilde{\bm{a}})), \ Q_n(\bm{T}^{-1}(\bm{\alpha}))=Q_n^{\mathcal{G}}(\bm{\alpha}).
	\label{eqn5}
	\end{equation}
	
	Considering the neighborhood of $\widetilde{\bm{a}}^0$ (true value of $\widetilde{\bm{a}}$),
	\begin{equation*}
	\Theta_n=\left\{\widetilde{\bm{a}}\in \mathbb{R}^{pnL}: \|\{\widetilde{\bm{a}}-\widetilde{\bm{a}}^0\|_\infty\leq r_n\right\}.
	\end{equation*}
	By Theorem 2, there is an event $E_1$ satisfying  $P(E_1)\geq  1- {2pL (\sum_{k=1}^p G_k^*)}/{n}$, and on the event $E_1$,
	\begin{equation}
	\|\hat{\widetilde{\bm{a}}}^\text{or}-\widetilde{\bm{a}}^0\|_\infty\leq r_n,
	\label{aridus}
	\end{equation}
	where $\hat{\widetilde{\bm{a}}}^\text{or}$ is the corresponding permutation of $\hat{{\bm{a}}}^\text{or}$, and $\hat{{\bm{a}}}^\text{or}$ can be found in (12) of the main paper. Accordingly, on the event $E_1$,  we have $\hat{\widetilde{\bm{a}}}^\text{or}$ $\in \Theta_n$. 
	
	By the following two steps, we show that, with probability approaching one, $\hat{\widetilde{\bm{a}}}^\text{or}$ is a strictly local minimizer of $Q_n(\widetilde{\bm{a}})$, i.e., the objective function (4) of the SCVC with  $\mathbb{E}_{\text{MST}}$ in the main paper. To prove this, we use $Q_n(\widetilde{\bm{a}}^*)$  to link $Q_n(\hat{\widetilde{\bm{a}}}^\text{or})$ with $Q_n(\widetilde{\bm{a}})$.

	\begin{enumerate}[label={(\roman*)}]

		\item 
		On the event $E_1$, for any $\widetilde{\bm{a}} \in \Theta_n$,
		\begin{equation*}\label{i}
		Q_n(\widetilde{\bm{a}}^*)> Q_n(\hat{\widetilde{\bm{a}}}^\text{or}),\quad \text{if}\quad  \widetilde{\bm{a}}^*\neq \hat{\widetilde{\bm{a}}}^{or}.
		\end{equation*}		
		\vspace{-1cm}
		\item There is an event $E_2$ such that $P(E_2)>1-2n^{-1}$. On the event $E_1\cap E_2$, there  exists $\mathcal{B}_n$, a neighborhood of $\hat{\widetilde{\bm{a}}}^{or}$, such that 
		\begin{equation*}\label{ii}
		Q_n(\widetilde{\bm{a}})\geq Q_n(\widetilde{\bm{a}}^*),\quad \forall \ \  \widetilde{\bm{a}}\in \Theta_n\cap \mathcal{B}_n.
		\end{equation*}
	\end{enumerate}
	\vspace{-0.4cm}
	
	By the above two steps, on the event $E_1\cap E_2$,  we have $Q_n(\widetilde{\bm{a}})>Q_n(\hat{\widetilde{\bm{a}}}^\text{or})$ for any $ \widetilde{\bm{a}}\in \Theta_n\cap \mathcal{B}_n$ and $ \widetilde{\bm{a}}\neq \hat{\widetilde{\bm{a}}}^\text{or}$.Hence, $\hat{\widetilde{\bm{a}}}^\text{or}$ is a strict local minimizer of  $Q_n(\widetilde{\bm{a}})$ on the event $E_1\cap E_2$ with $P(E_1\cap E_2)\geq 1- {2pL (\sum_{k=1}^p G_k^*)}/{n}-2n^{-1}$.
	
	Now we prove the result in \ref{i}. We first show that $P_n^{\mathcal{G}}(\bm{T}^*{(\widetilde{\bm{a}})})=C_n$ for any $\widetilde{\bm{a}}\in \Theta_n$, where $C_n$ is a constant independent of $\widetilde{\bm{a}}$,  $\bm{T}^*{(\widetilde{\bm{a}})}=(\bm{T}^*_1(\widetilde{\bm{a}}_1)^T, \cdots, \bm{{T}}^*_p(\widetilde{\bm{a}}_p)^T)^T$, and $\bm{T}_k^*(\widetilde{\bm{a}}_{k})=(T_k^*(\widetilde{\bm{a}}_{k, 1})^T, \cdots, T_k^*(\widetilde{\bm{a}}_{k, L})^T)^T$, $k=1, \cdots, p$. For $\bm{\alpha}\in\left\{\bm{T}^*{(\widetilde{\bm{a}})}: \widetilde{\bm{a}}\in \Theta_n\right\}$,  by the definition of $P_n^{\mathcal{G}}({\bm{\alpha}})$ and Assumption \ref{Au4},  {to prove $P_n^{\mathcal{G}}(\bm{T}^*{(\widetilde{\bm{a}})})=C_n$,}  it is sufficient to prove $\|\widetilde{\bm{\alpha}}_{k, g_k^*}-\widetilde{\bm{\alpha}}_{k, {g_k^*}'}\|_2> a_0\lambda_k$ for any $1\leq g_k^*< {g_k^*}'\leq G_k^*$ and $|\mathbb{C}_{g_k^*{g_k^*}'}|\neq 0$, { where the positive constant $a_0$ can be found in Assumption \ref{Au4}}}. Noticing 
\begin{equation}
\begin{split}
\|\widetilde{\bm{\alpha}}_{k, g_k^*}-\widetilde{\bm{\alpha}}_{k, {g_k^*}'}\|_2
&\geq L^{-1/2}\|\widetilde{\bm{\alpha}}_{k, g_k^*}-\widetilde{\bm{\alpha}}_{k, {g_k^*}'}\|_1\\
&\geq L^{-1/2}(\|\widetilde{\bm{\alpha}}_{k, g_k^*}^0-\widetilde{\bm{\alpha}}_{k, {g_k^*}'}^0\|_1-2L \|{\bm{\alpha}}-{\bm{\alpha}}^0\|_\infty)\\
&\geq L^{-1/2}(\|\widetilde{\bm{\alpha}}_{k, g_k^*}^0-\widetilde{\bm{\alpha}}_{k, {g_k^*}'}^0\|_2-2L \|{\bm{\alpha}}-{\bm{\alpha}}^0\|_\infty),
\label{eqn6}
\end{split}
\end{equation}
and
\begin{equation}\label{eqn7}
\begin{split}
\|{\bm{\alpha}}-{\bm{\alpha}}^0\|_\infty&=\underset{
	\begin{scriptsize}
	\begin{split}
	\vspace{-1cm}	 &1\leq k\leq p, 1\leq l\leq L,\\
	&\quad \ 1\leq g_k^*\leq G_k^*
	\end{split}
	\end{scriptsize}
}{\text{max}}
|\widetilde{\alpha}_{k,g_k^*}^l-\widetilde{\alpha}_{k,g_k^* }^{l, 0}|=\underset{
	\begin{scriptsize}
	\begin{split}
	\vspace{-1cm}	 &1\leq k\leq p, 1\leq l\leq L,\\
	&\quad \ 1\leq g_k^*\leq G_k^*
	\end{split}
	\end{scriptsize}
}{\text{max}}\Big|\sum_{i\in \mathcal{M}^{{g_k^*}}_{k}} \frac{a_{k, i}^l-a_{k, i }^{l, 0}}{|\mathcal{M}^{{g_k^*}}_{k}|}\Big|\\
&\leq \underset{
	\begin{scriptsize}
	\begin{split}
	\vspace{-1cm}	 &1\leq k\leq p, 1\leq l\leq L,\\
	&\quad \ 1\leq g_k^*\leq G_k^*
	\end{split}
	\end{scriptsize}
}{\text{max}}\{ \underset{i\in \mathcal{M}^{{g_k^*}}_{k}}{\text{max}}|{a_{k, i}^l-a_{k, i}^{l, 0}}|\}=\|\widetilde{\bm{a}}-\widetilde{\bm{a}}^0\|_\infty\leq r_n,
\end{split}
\end{equation}
so that $\|\widetilde{\bm{\alpha}}_{k, g_k^*}-\widetilde{\bm{\alpha}}_{k, {g_k^*}'}\|_2\geq L^{-1/2}(\vartheta_{k}-2Lr_n) >a_0 \lambda_k$,  {following the assumption $\vartheta_{k}>> 2\sqrt{L} a_0\lambda_k>>r_n$.} Thus, for any $\widetilde{\bm{a}}\in \Theta_n$, we have $P_n^{\mathcal{G}}(\bm{T}^*{(\widetilde{\bm{a}})})=C_n$, so   $Q_n^{\mathcal{G}}(\bm{T}^*{(\widetilde{\bm{a}})})=L_n^{\mathcal{G}}(\bm{T}^*{(\widetilde{\bm{a}})})+C_n$. {Moreover, on the event $E_1$, we have $\hat{\widetilde{\bm{a}}}^\text{or}\in \Theta_n$  by (\ref{aridus}), so $P_n^{\mathcal{G}}(\hat{\bm{\alpha}}^\text{or})=C_n$.} Since $ \hat{\bm{\alpha}}^\text{or}$ is the unique global minimizer of $L_n^{\mathcal{G}}({\bm{\alpha}})$, then $L_n^{\mathcal{G}}(\bm{T}^*{(\widetilde{\bm{a}})})>L_n^{\mathcal{G}}(\hat{\bm{\alpha}}^\text{or})$ for $\bm{T}^*{(\widetilde{\bm{a}})}\neq \hat{\bm{\alpha}}^\text{or}$.  Thus,  on the event $E_1$, $Q_n^{\mathcal{G}}(\bm{T}^*{(\widetilde{\bm{a}})})>Q_n^{\mathcal{G}}(\hat{\bm{\alpha}}^\text{or})$. Following (\ref{eqn5}),  it is straightforward to obtain $	Q_n(\widetilde{\bm{a}}^*)> Q_n(\hat{\widetilde{\bm{a}}}^\text{or}) $ for $ \widetilde{\bm{a}}^*\neq \hat{\widetilde{\bm{a}}}^\text{or}$. The result in \ref{i} is proved.

Then, we prove \ref{ii}. Let
\begin{equation*}
\mathcal{B}_n=\left\{\widetilde{\bm{a}}\in \mathbb{R}^{pnL}: \|\widetilde{\bm{a}}-\hat{\widetilde{\bm{a}}}^\text{or}\|_2\leq t_n\right\},
\end{equation*}
where $\{t_n\}$ is a positive sequence. 
For $ \widetilde{\bm{a}}\in \Theta_n\cap \mathcal{B}_n$, by Taylor expansion, we have
\begin{equation*}
Q_n(\widetilde{\bm{a}})- Q_n(\widetilde{\bm{a}}^*)=I_3+I_4,
\end{equation*}
where
\begin{equation*}
I_3=-\frac{1}{n}\sum_{k=1}^p\sum_{l=1}^L \left(\bm{Y}-\sum_{k=1}^p\sum_{l=1}^L\widetilde{\mathbb{Z}}_l^k\widetilde{\bm{a}}_{k, l}^m\right)^T(\widetilde{\mathbb{Z}}_l^k)^T (\widetilde{\bm{a}}_{k, l}-\widetilde{\bm{a}}_{k,l}^*)+2 \sum_{k=1}^p \varrho_k \sum_{l=1}^L h_{ll} (\widetilde{\bm{a}}_{k, l}^m)^T (\widetilde{\bm{a}}_{k, l}-\widetilde{\bm{a}}_{k, l}^*),
\end{equation*}
and
\begin{equation*}
I_4=\sum_{k=1}^p\sum_{i=1}^n \frac{\partial P_n(\widetilde{\bm{a}}^m)}{\partial \bm{a}_{k, i}} (\bm{a}_{k, i}-\bm{a}_{k,i}^*),
\end{equation*}
where $\widetilde{\bm{a}}_{k, l}^m=\pi \widetilde{\bm{a}}_{k, l}+(1-\pi)\widetilde{\bm{a}}^*_{k, l}$, and $\widetilde{\bm{a}}^m=\pi \widetilde{\bm{a}}+(1-\pi)\widetilde{\bm{a}}^*$ for some $\pi\in (0,1)$. 

For $I_4$, by the definition of $P_n(\widetilde{\bm{a}}^m)$,  we have
\begin{equation*}
I_4=\sum_{k=1}^p\lambda_k{\sum_{(i,j)\in \mathbb{E}_\text{MST}}}\bar{\bm{\rho}}_k(\|\bm{a}_{k, i}^m-\bm{a}_{k, j}^m\|_2)^T\{(\bm{a}_{k, i}-\bm{a}_{k, i}^*)-(\bm{a}_{k, j}-\bm{a}_{k, j}^*)\}, 
\end{equation*}
where $\bar{\bm{\rho}}_k(\|\bm{a}_{k, i}^m-\bm{a}_{k, j}^m\|_2)=\left(\bar{\rho}_{k, 1}(\|\bm{a}_{k, i}^m-\bm{a}_{k, j}^m\|_2), \cdots, \bar{\rho}_{k, L}(\|\bm{a}_{k, i}^m-\bm{a}_{k, j}^m\|_2)\right)^T$,  { the definition of $\bar{\rho}_{k, l}(\cdot),\ l=1, \cdots, L$, is given in the beginning of Section \ref{Sect1}}, and $\bm{a}_{k, i}^m=\pi\bm{a}_{k, i}+(1-\pi)\bm{a}_{k, i}^*$, $i=1, \cdots, n$. When $i, j\in \mathcal{M}^{{g_k^*}}_{k}$ for any $g_k^*\in\{1, \cdots, G_k^*\}$, we know $\bm{a}_{k, i}^*=\bm{a}_{k, j}^*$, hence  $\bm{a}_{k, i}^m-\bm{a}_{k, j}^m= \pi(\bm{a}_{k, i}-\bm{a}_{k, j})$. Then, 
\begin{equation*}
\begin{split}
I_4=&\sum_{k=1}^p\lambda_k\sum_{g_k^*=1}^{G_k^*}{\sum_{(i,j)\in \mathbb{E}_\text{MST},\ i,j \in \mathcal{M}^{{g_k^*}}_{k} }}{{\rho}}'_k(\|\bm{a}_{k, i}^m-\bm{a}_{k, j}^m\|_2)\|\bm{a}_{k, i}-\bm{a}_{k, j}\|_2\\
&+\sum_{k=1}^p \lambda_k
\underset{
	\begin{scriptsize}
	\begin{split}
	\vspace{-1cm}	 &i,j\notin \mathcal{M}^{{g_k^*}}_{k}, g_k^*=1, \cdots, G_k^*,\\
	&\quad\quad\quad\quad (i,j)\in \mathbb{E}_\text{MST}
	\end{split}
	\end{scriptsize}
}{\sum}
\bar{\bm{\rho}}_k(\|\bm{a}_{k, i}^m-\bm{a}_{k, j}^m\|_2)^T\{(\bm{a}_{k, i}-\bm{a}_{k, i}^*)-(\bm{a}_{k, j}-\bm{a}_{k, j}^*)\}.
\end{split}
\end{equation*}
Similar with the proving process in (\ref{eqn6}),  for $(i,j)\in \mathbb{E}_\text{MST},\ i,j\notin \mathcal{M}^{{g_k^*}}_{k}, g_k^*=1, \cdots, G_k^*$, $k=1, \cdots, p$, we can obtain $\|\bm{a}_{k, i}^m-\bm{a}_{k, j}^m\|_2>a_0\lambda_k$, thus $\bar{\bm{\rho}}_k(\|\bm{a}_{k, i}^m-\bm{a}_{k, j}^m\|_2)=\bm{0}$ by Assumption \ref{Au4}. Hence,
\begin{equation}
\begin{split}
I_4=\sum_{k=1}^p\lambda_k\sum_{g_k^*=1}^{G_k^*}{\sum_{(i,j)\in \mathbb{E}_\text{MST},\ i,j \in \mathcal{M}^{{g_k^*}}_{k} }}{{\rho}}'_k(\|\bm{a}_{k, i}^m-\bm{a}_{k, j}^m\|_2)\|\bm{a}_{k, i}-\bm{a}_{k, j}\|_2.
\label{eqn8}
\end{split}
\end{equation}
Similar with the proof of (\ref{eqn7}), we have $\|\bm{a}^*-\hat{\bm{a}}^\text{or}\|_\infty\leq \|\bm{a}-\hat{\bm{a}}^\text{or}\|_\infty$. Then, for $(i,j)\in \mathbb{E}_\text{MST},\ i,j \in \mathcal{M}^{{g_k^*}}_{k}$,
\begin{equation*}
\begin{split}
\|\bm{a}_{k, i}^m-\bm{a}_{k, j}^m\|_2&\leq \|\bm{a}_{k, i}^m-\bm{a}_{k, j}^m\|_1\leq 2L \|\bm{a}^m-\bm{a}^*\|_\infty\leq 2L \|\bm{a}-\bm{a}^*\|_\infty\\
&\leq 4L \|\bm{a}-\hat{\bm{a}}^\text{or}\|_\infty\leq 4Lt_n.
\end{split}
\end{equation*}
Therefore, ${{\rho}}'_k(\|\bm{a}_{k, i}^m-\bm{a}_{k, j}^m\|_2)\geq {{\rho}}'_k(4Lt_n)$ by the concavity of $\rho_k(\cdot)$. According to (\ref{eqn8}), 
\begin{equation*}
\begin{split}
I_4\geq \sum_{k=1}^p\lambda_k{{\rho}}'_k(4Lt_n)\sum_{g_k^*=1}^{G_k^*}{\sum_{(i,j)\in \mathbb{E}_\text{MST},\ i,j \in \mathcal{M}^{{g_k^*}}_{k} }}\|\bm{a}_{k, i}-\bm{a}_{k, j}\|_2.
\end{split}
\end{equation*}
For every $i^*, j^*\in  \mathcal{M}^{{g_k^*}}_{k}$, according to the definition of $ \mathcal{M}^{{g_k^*}}_{k}$, we know there is  a path $i^*=p_1 \to p_2\to\cdots \to p_M=j^*,  $ connecting $i^*$ and $j^*$, where $(p_{m-1}, p_m)\in \widetilde{\mathbb{E}}^{g_k^*}_k$, $m=2, \cdots, M$,  and $\widetilde{\mathbb{E}}^{g_k^*}_k=\{(i, j):(i, j)\in \mathbb{E}_{\text{MST}},\ \text{and}\  i, j\in  \mathcal{M}^{{g_k^*}}_{k}\}$. Moreover,  we can always let  the path satisfy that $\{p_m\}_{1}^M$ are mutually unequal, otherwise suppose $p_{m_1}=p_{m_2}$, $m_1<m_2$, the path after deleting $p_{m_1+1}\to \cdots \to p_{m_2}$ still can connect $i^* ,j^*$. Therefore,  for every $i^*, j^*\in \mathcal{M}^{{g_k^*}}_{k}$,
\begin{equation*}
\begin{split}
\|\bm{a}_{k, i^*}&-\bm{a}_{k, j^*}\|_2\leq {\sum_{(i,j)\in \mathbb{E}_\text{MST},\ i,j \in \mathcal{M}^{{g_k^*}}_{k} }}\|\bm{a}_{k, i}-\bm{a}_{k, j}\|_2 \\
&\Longrightarrow {\sum_{i<j,\ i,j \in \mathcal{M}^{{g_k^*}}_{k}}}\|\bm{a}_{k, i}-\bm{a}_{k, j}\|_2\leq \frac{|\mathcal{M}^{{g_k^*}}_{k}|(|\mathcal{M}^{{g_k^*}}_{k}|-1)}{2} {\sum_{(i,j)\in \mathbb{E}_\text{MST},\ i,j \in\mathcal{M}^{{g_k^*}}_{k}}}\|\bm{a}_{k, i}-\bm{a}_{k, j}\|_2.
\end{split}
\end{equation*}
Hence,
\begin{equation}
\begin{split}
I_4\geq \sum_{k=1}^p \lambda_k{{\rho}}'_k(4Lt_n)\sum_{g_k^*=1}^{G_k^*}  \frac{2}{|\mathcal{M}^{{g_k^*}}_{k}|(|\mathcal{M}^{{g_k^*}}_{k}|-1)}   {\sum_{i<j,\ i,j \in \mathcal{M}^{{g_k^*}}_{k} }}\|\bm{a}_{k, i}-\bm{a}_{k, j}\|_2.
\label{eqnnn9}
\end{split}
\end{equation}

Now we consider $I_3$. Denote $\bm{W}_l^k=(w_{1, l}^k, \cdots, w_{n, l}^k)^T=-\widetilde{\mathbb{Z}}_l^k\left(\bm{Y}-\sum_{k=1}^p\sum_{l=1}^L\widetilde{\mathbb{Z}}_l^k\widetilde{\bm{a}}_{k, l}^m\right)+2n\varrho_k h_{ll} \widetilde{\bm{a}}_{k, l}^m $. Then, we have
\begin{equation*}
\begin{split}
I_3
&=\frac{1}{n} \sum_{k=1}^p \sum_{l=1}^L (\bm{W}_l^k)^T (\widetilde{\bm{a}}_{k, l}-\widetilde{\bm{a}}_{k, l}^*)=\frac{1}{n}  \sum_{k=1}^p \sum_{l=1}^L \sum_{g_k^*=1}^{G_k^*}\sum_{i\in \mathcal{M}^{{g_k^*}}_{k}} w_{i, l}^k ({{a}}_{k, i}^l-{{a}}_{k, i}^{l, *})\\
&=\frac{1}{n}  \sum_{k=1}^p \sum_{l=1}^L \sum_{g_k^*=1}^{G_k^*}\sum_{i\in \mathcal{M}^{{g_k^*}}_{k}} w_{i, l}^k \Big({{a}}_{k, i}^l-\frac{1}{|\mathcal{M}^{{g_k^*}}_{k}|}\sum_{j \in \mathcal{M}^{{g_k^*}}_{k}}{{a}}_{k, j}^l\Big)\\
&=\frac{1}{n}\sum_{k=1}^p \sum_{l=1}^L \sum_{g_k^*=1}^{G_k^*}\sum_{i, j\in \mathcal{M}^{{g_k^*}}_{k}}  \frac{w_{i, l}^k}{|\mathcal{M}^{{g_k^*}}_{k}|} \left({{a}}_{k, i}^l-{{a}}_{k, j}^l\right)\\
&=\frac{1}{n} \sum_{k=1}^p \sum_{l=1}^L \sum_{g_k^*=1}^{G_k^*}\sum_{i<j,\ i, j\in \mathcal{M}^{{g_k^*}}_{k}} \frac{w_{i, l}^k}{|\mathcal{M}^{{g_k^*}}_{k}|} \left({{a}}_{k, i}^l-{{a}}_{k, j}^l\right)+\frac{1}{n} \sum_{k=1}^p \sum_{l=1}^L \sum_{g_k^*=1}^{G_k^*}\sum_{i<j,\ i, j\in \mathcal{M}^{{g_k^*}}_{k}} \frac{w_{j, l}^k}{|\mathcal{M}^{{g_k^*}}_{k}|} \left({{a}}_{k, j}^l-{{a}}_{k, i}^l\right)\\
&=\frac{1}{n} \sum_{k=1}^p \sum_{l=1}^L \sum_{g_k^*=1}^{G_k^*}\sum_{i<j,\ i, j\in \mathcal{M}^{{g_k^*}}_{k}} \frac{w_{i, l}^k-w_{j, l}^k}{|\mathcal{M}^{{g_k^*}}_{k}|} \left({{a}}_{k, i}^l-{{a}}_{k, j}^l\right).\\
\end{split}
\end{equation*}
Hence,
\begin{equation}
\begin{split}
|I_3|\leq \frac{1}{n} \underset{
	\begin{scriptsize}
	\begin{split}
	\vspace{-1cm}	 &1\leq k\leq p, 1\leq l\leq L\\
	&\quad \quad 1\leq i, j\leq n
	\end{split}
	\end{scriptsize}
}{\text{max}}
|w_{i,l}^k-w_{j,l}^k|\sum_{k=1}^p\sum_{g_k^*=1}^{G_k^*}\frac{\sqrt{L}}{|\mathcal{M}^{{g_k^*}}_{k}|}\sum_{i<j,\ i, j\in \mathcal{M}^{{g_k^*}}_{k}} \|\bm{a}_{k, i}-\bm{a}_{k, j}\|_2.
\label{eqnn9}
\end{split}
\end{equation}
Moreover, for any $k\in \{1, \cdots, p\}$ and $l\in \{1, \cdots, L\}$, 
\begin{equation}
\begin{split}
\underset{1\leq i, j\leq n}{\text{max}} &|w_{i,l}^k-w_{j,l}^k|\leq 2\|\bm{W}_l^k\|_\infty \\
&\leq 2 \|\widetilde{\mathbb{Z}}_l^k\|_\infty\Big\{\|\bm{\epsilon}\|_\infty+\sum_{k=1}^p \sum_{l=1}^L\|\widetilde{\mathbb{Z}}_l^k\|_\infty \|\widetilde{\bm{a}}_{k, l}^m-\widetilde{\bm{a}}_{k, l}^0\|_\infty\Big\}+2n\varrho_k \|\widetilde{\bm{a}}_{k, l}^m\|_\infty.
\label{eqn9}
\end{split}
\end{equation}
Similar with the proof of (\ref{eqn7}), we have 
$\|\widetilde{\bm{a}}_{k, l}^*-\widetilde{\bm{a}}_{k, l}^0\|_\infty\leq \|\widetilde{\bm{a}}_{k, l}-\widetilde{\bm{a}}_{k, l}^0\|_\infty$, so that
\begin{equation}
\|\widetilde{\bm{a}}_{k, l}^m-\widetilde{\bm{a}}_{k, l}^0\|_\infty\leq \pi \|\widetilde{\bm{a}}_{k, l}-\widetilde{\bm{a}}_{k, l}^0\|_\infty+(1-\pi)\|\widetilde{\bm{a}}_{k, l}^*-\widetilde{\bm{a}}_{k, l}^0\|_\infty\leq \|\widetilde{\bm{a}}-\widetilde{\bm{a}}^0\|_\infty \leq r_n.
\end{equation}
Then,
\begin{equation}
\|\widetilde{\bm{a}}_{k, l}^m\|_\infty\leq \|\widetilde{\bm{a}}_{k, l}^0\|_\infty+r_n\leq \|{\bm{a}}^0\|_\infty+r_n.
\end{equation}
By Assumption \ref{Au6},
\begin{equation}
P(\|\bm{\epsilon}\|_\infty>\sqrt{2c_1^{-1}}\sqrt{\text{log}\ n})\leq 2n^{-1}.
\label{sseqn12}
\end{equation}
Thus,  there is an event $E_2$ such that $P(E_2^c)\leq 2n^{-1}$, and on the event $E_1\cap E_2$,  by (\ref{eqn9})-(\ref{sseqn12}) and {Assumption \ref{Au5}, }  we have
\begin{equation*}
\begin{split}
\underset{
	\begin{scriptsize}
	\begin{split}
	\vspace{-1cm}	 &1\leq k\leq p, 1\leq l\leq L\\
	&\quad \quad 1\leq i, j\leq n
	\end{split}
	\end{scriptsize}
}{\text{max}}|w_{i, l}^k-w_{j, l}^k|
\leq 2 C_3\left\{\sqrt{2c_1^{-1}}\sqrt{\text{log}\ n}+ pLC_3 r_n\right\}+2n\widetilde{\varrho} (\|{\bm{a}}^0\|_\infty+r_n)=\psi_n.
\end{split}
\end{equation*}
Combining (\ref{eqnn9}), we have 
\begin{equation}
|I_3|\leq \frac{\psi_n}{n}\sum_{k=1}^p\sum_{g_k^*=1}^{G_k^*}\frac{\sqrt{L}}{|\mathcal{M}^{{g_k^*}}_{k}|}\sum_{i<j,\ i, j\in \mathcal{M}^{{g_k^*}}_{k}} \|\bm{a}_{k, i}-\bm{a}_{k, j}\|_2.
\label{sseqn15}
\end{equation}
Let $t_n=o(1)$, then ${{\rho}}_k'(4Lt_n)\to 1$. Thus, by (\ref{eqnnn9}) and (\ref{sseqn15}), under the rate assumption of $\lambda_k$ and $\varrho_k$ in Theorem 3, we have
\begin{equation*}
\begin{split}
&Q_n(\widetilde{\bm{a}})- Q_n(\widetilde{\bm{a}}^*)=I_3+I_4\\
&\geq \sum_{k=1}^p\sum_{g_k^*=1}^{G_k^*}  \left\{\frac{2\lambda_k{{\rho}}'_k(4Lt_n)}{|\mathcal{M}^{{g_k^*}}_{k}|(|\mathcal{M}^{{g_k^*}}_{k}|-1)}- \frac{\psi_n\sqrt{L}}{n|\mathcal{M}^{{g_k^*}}_{k}|}\right\} {\sum_{i<j,\ i,j \in \mathcal{M}^{{g_k^*}}_{k} }}\|\bm{a}_{k, i}-\bm{a}_{k, j}\|_2\geq 0.
\end{split}
\end{equation*}
The result in \ref{ii} is proved. Together with the result in \ref{i}, proof is completed.  $\quad\square$

\section{The Nelder–Mead algorithm}
Following  \cite{singer2009nelder}, the Nelder–Mead algorithm for minimizing $\text{BIC}$ $(\{\lambda_k, \varrho_k\}_{k=1}^p)$ contains the following steps. 
\begin{enumerate}[itemindent=3.5em, label={Step \arabic*.}]
	\item {\bf{Initial simplex.}}
	
	The initial simplex $S$ is  constructed by generating $2p+1$ vertices, i.e., $\bm{x}_j=(\lambda_1^j, \cdots, \lambda_p^j, \varrho_1^j, \cdots, \varrho_p^j)$, $j=0, \cdots, 2p.$ The common method for generating $\bm{x}_j$ is
	\begin{equation*}
	\bm{x}_j=\bm{x}_0+h_j \bm{e}_j, \quad j=1, \cdots, 2p,
	\end{equation*}
	where $h_j\in \mathbb{R}$ is the step size, and $\bm{e}_j\in \mathbb{R}^{2p}$ is the unit vector with $j-$th element equal to one, others equal to zero.  A decent choice of $\bm{x}_0$ and $h_j$ can be obtained by comparing the value of objective function over a small number of grid. To be specific, we calculate the BIC values over a small number of grid, and take the $\bm{x}_0$ as the point corresponding to the smallest BIC value, and $\bm{h}=(h_1, \cdots, h_{2p})^T$ can take the value with magnitude being the same as  $\bm{x}_0$, such as $\bm{h}=\bm{x}_0/2$. 
	
	\item {\bf{Ordering.}}
	
	Order according to the values at these vertices:
	\begin{equation*}
	\text{BIC}(\bm{x}_{(0)})\leq \text{BIC}(\bm{x}_{(1)}) \leq \cdots \leq  \text{BIC}(\bm{x}_{(2p)}),
	\end{equation*}
	{where $\bm{x}_{(j)}$ is the corresponding vertex with $(j+1)$-th smallest BIC value.}
	If $\bm{x}_{(j)}$, $j=0, \cdots, 2p$ are close to each other, terminate the algorithm, and take $\bm{x}_{(0)}$ as the minimizer of the objective function. If not, let $\bm{x}_c=\frac{1}{2p} \sum_{j=0}^{2p-1} \bm{x}_{(j)}$.
	
	\item {\bf{Reflection.}}
	
	Compute reflected point $\bm{x}_r=\bm{x}_c+\alpha(\bm{x}_c-\bm{x}_{(2p)})$ with $\alpha>0$. If the reflected point satisfies $\text{BIC}(\bm{x}_{(0)})\leq \text{BIC}(\bm{x}_r) < \text{BIC}(\bm{x}_{(2p-1)})$, then obtain a new simplex by replacing the worst point $\bm{x}_{(2p)}$ with the reflected point $\bm{x}_r$, and go to Step 2.
	
	\item {\bf{Expansion.}}
	
	If the reflected point satisfies $\text{BIC}(\bm{x}_r)<\text{BIC}(\bm{x}_{(0)})$, then compute the expanded point $\bm{x}_e=\bm{x}_c+\gamma (\bm{x}_r-\bm{x}_c)$ with $\gamma>1$. If $\text{BIC}(\bm{x}_e)<\text{BIC}(\bm{x}_{r})$, then obtain a new simplex by replacing the worst point  $\bm{x}_{(2p)}$ with the expanded point $\bm{x}_e$, and go to Step 2, else obtain  a new simplex by replacing the worst point  $\bm{x}_{(2p)}$ with the reflected point $\bm{x}_r$, and go to Step 2.
	
	\item {\bf{Contraction.}}
	
	If the reflected point satisfies $\text{BIC}(\bm{x}_r)\geq\text{BIC}(\bm{x}_{(2p-1)})$, then compute the contracted point $\bm{x}_t=\bm{x}_c+\rho (\bm{x}_{(2p)}-\bm{x}_c)$ with $0<\rho\leq 0.5$. If the contracted point satisfies $\text{BIC}(\bm{x}_t)<\text{BIC}(\bm{x}_{(2p)})$, then obtain a new simplex by replacing the worst point $\bm{x}_{(2p)}$ with the contracted  point $\bm{x}_t$, and go to Step 2.
	\item {\bf{Shrink.}}
	
	If the contracted point satisfies $\text{BIC}(\bm{x}_t)\geq\text{BIC}(\bm{x}_{(2p)})$, 
	then obtain a new simplex by replacing
	$\bm{x}_{(j)}$ with $\bm{x}_{(0)}+\sigma(\bm{x}_{(j)}-\bm{x}_{(0)})$, $j=1, \cdots, 2p$, and go to Step 2.
	
\end{enumerate}

The standard values, used in most implementations are $\alpha=1$, $\gamma=2$, $\rho=0.5$ and $\sigma=0.5.$

{\vspace{-0.7cm}
	\section{Generate spatially clustered pattern}
	Here, we take the spatially clustered patterns of $\beta_2(\bm{s})$ in  Figure 2(a) of the main paper as an example, to demonstrate how to generate spatially clustered patterns of MST-equal and MST-unequal.
	
	For the MST-equal pattern of $\beta_2(\bm{s})$ in  Figure 2(a), we construct it through following steps,
	\begin{enumerate}[itemindent=3.5em, label={Step \arabic*.}]

		\item Randomly generate a location $({s}_1, {s}_2)$, where $s_1$ and $s_2$ are from [0, 1] uniform distribution.
		
		\item Compute the distance from $({s}_1, {s}_2)$ to three lines $y=x+0.5$, $y=x$ and $y=x-0.5$, respectively. Denote them as $d_1, d_2$ and $d_3$.
		
		\item Set a tolerance parameter $\delta=0.02$, if $\text{min}(d_1, d_2, d_3)\geq\delta$, keep this location, otherwise abandon it.
		
		\item Repeat Step 1-3 until the number of locations reaches 1000, then we form four clusters based on these 1000 locations, which is defined by $\{\bm{s}_i: s_2>s_1+0.5\}$, $\{\bm{s}_i: s_1+0.5\geq s_2>s_1\}$,   $\{\bm{s}_i:  s_1 \geq s_2> s_1-0.5\}$ and  $\{\bm{s}_i:  s_1-0.5 \geq s_2 \}$. And the values of $\beta_2(\bm{s})$ in these four clusters are 1, -1, 0.5, -0.5, respectively.
		
	\end{enumerate}
	
	{
		The tolerance parameter $\delta$ controls the minimum distance between different proximate clusters, which can be easily observed in Figure 2 of the main paper.  As discussed in Section 2.3.3 of the main paper, MST only connects the proximate locations. Thus, if $\delta$ is relatively large, i.e., the minimum distance between different proximate clusters is relatively large, all the locations within the same cluster are more likely to be connected through the edges of MST, resulting in the MST-equal pattern; if $\delta$ is relatively small, some locations may be isolated from its belonging cluster and connected to a different cluster due to  closer distance.  
		In the Step 3 above, we set the tolerance parameter $\delta=0.02$ to generate the MST-equal pattern. To generate the MST-unequal pattern, we set  $\delta=0.01$ and details are given as follows. 
	}
	
	For the MST-unequal pattern of $\beta_2(\bm{s})$ in  Figure 2(a),   we construct it through following steps,
	\begin{enumerate}[itemindent=3.5em, label={Step* \arabic*.}]

		\item The same as Step 1.
		
		\item The same as Step 2.
		
		\item Set a tolerance parameter $\delta=0.01$, if $\text{min}(d_1, d_2, d_3)\geq\delta$, keep this location, otherwise abandon it.
		
		\item Repeat Step* 1-3 until the number of locations reaches 1000.
		We form four clusters of $\beta_2(\bm{s})$  by following two steps. First, divide these 1000 locations into four parts, that is,   $\{\bm{s}_i: s_2>s_1+0.5\}$, $\{\bm{s}_i: s_1+0.5\geq s_2>s_1\}$,   $\{\bm{s}_i:  s_1 \geq s_2> s_1-0.5\}$ and  $\{\bm{s}_i:  s_1-0.5 \geq s_2 \}$, and the values of $\beta_2(\bm{s})$ in these four parts are 1, -1, 0.5, -0.5, respectively. Second,  based on these four parts, we can  obtain the corresponding $\mathcal{M}^{{g_k^*}}_k$, $g_k^*=1, \cdots, G_k^*$. For some $g_{k,1}^*\in\{1, \cdots, G_k^*\}$,
		the sample size in $\mathcal{M}^{g_{k,1}^*}_{k}$ may be one or two, which violates the theoretical requirement of the sample size in  $\mathcal{M}^{g_{k,1}^*}_{k}$. Thus, we form  four clusters ( locations in the same cluster have the same value of $\beta_2(\bm{s})$),  through replacing the value of $\beta_2(\bm{s}_i)$,  $i\in \mathcal{M}^{g_{k,1}^*}_{k}$ with the value of   $\beta_2(\bm{s}_j)$, $j \in \mathcal{M}^{g_{k,2}^*}_{k}$, $g_{k,2}^*\in\{1, \cdots, G_k^*\}$ , where the sample size of $\mathcal{M}^{g_{k,2}^*}_{k}$ is relatively large, and  $\mathcal{M}^{g_{k,1}^*}_{k}$ and $\mathcal{M}^{g_{k,2}^*}_{k}$ are connected by the edge of MST.
	\end{enumerate}
}
\vspace{-0.7cm}
\section{Simulation study: SCC* with the initial values set as the true values}
{The setting of this simulation study is the same as that in Section 4.1 of the main paper.  
	Table \ref{ttable1} summarizes the results of the {\it infeasible}  SCC* with the initial values set as the true values. Compared to the results of Table 1 in the main paper,  the {\it infeasible}  SCC*  gives uniformly better performance than SCVC, SCC, GWR, PSE, and SCC* with the initial values set by the SCC estimates. It is because the SCC model is the true model and the initial values are set as the true values.}

\begin{table}[htbp] 
	\hspace{0.6cm}
	\small
	\renewcommand\arraystretch{0.8}
	\setlength{\abovecaptionskip}{0pt}
	\setlength{\belowcaptionskip}{10pt}
	\hskip-1.0cm
	\begin{threeparttable}[b]
		\caption{The summarized results of SCC* under Study 1 in the main paper (the SCC model is the true model in Study 1), with the initial values set as the true values. } 
		\begin{tabular}{p{2.4cm}<{\centering}p{1.6cm}<{\centering}p{1.5cm}<{\centering}p{1.5cm}<{\centering}p{1.3cm}<{\centering}p{1.3cm}<{\centering}p{1cm}<{\centering}p{1cm}<{\centering}} 
			\hline
			\hline
			\multicolumn{1}{p{2.4cm}<{\centering}}{ Pattern}&\multicolumn{1}{p{1cm}<{\centering}}{Correlation} &  \multicolumn{1}{p{1.5cm}<{\centering}}{$\text{MSE}_{\beta_2}$}&\multicolumn{1}{p{1.5cm}<{\centering}}{$\text{MSE}_{\beta_1}$}&\multicolumn{1}{p{1.25cm}<{\centering}}{$\text{RI}_2$}&\multicolumn{1}{p{1.25cm}<{\centering}}{$\text{RI}_1$}&\multicolumn{1}{p{1.25cm}<{\centering}}{$\text{IC}_2$} &\multicolumn{1}{p{1.25cm}<{\centering}}{$\text{IC}_1$}\\
			\cmidrule(lr){3-4}\cmidrule(lr){5-6}\cmidrule(lr){7-8}
			&weak&0.0004&0.0004&100.00&100.00&4&4\\
			MST-equal&&(0.0000)&(0.0000)&(0.00)&(0.00)&(0.00)&(0.00)\\
			&strong&0.0008&0.0008&100.00&100.00&4&4\\
			&&(0.0001)&(0.0001)&(0.00)&(0.00)&(0.00)&(0.00)\\
			\hline
			&weak&0.0008&0.0012&85.75&88.18&7&8\\
			MST-unequal&&(0.0001)&(0.0001)&(0.00)&(0.00)&(0.00)&(0.00)\\
			&strong&0.0018&0.0041&85.75&88.18&7&8\\
			&&(0.0001)&(0.0001)&(0.00)&(0.00)&(0.00)&(0.00)\\
			\hline
			\hline
		\end{tabular}
		\label{ttable1}
		\begin{tablenotes} 
			\item SCC*: spatially clustered coefficient regression based on SCAD. $\text{MSE}_{\beta_k}$: mean squared error $(\times 10)$ for $k$-th covariate, $k=1, 2$;  $\text{RI}_k$: rand index $(\times 100)$ for $k$-th covariate; $\text{IC}_k$: the number of identified clusters for $k$-th covariate.  Values in the parentheses are the standard errors. 
		\end{tablenotes}
	\end{threeparttable}
\end{table}

\vspace{-0.5cm}
\section{Simulation study: Smooth-varying coefficients}
\begin{figure}[htbp]
	\centering	
	\includegraphics[width=6in,height=3in]{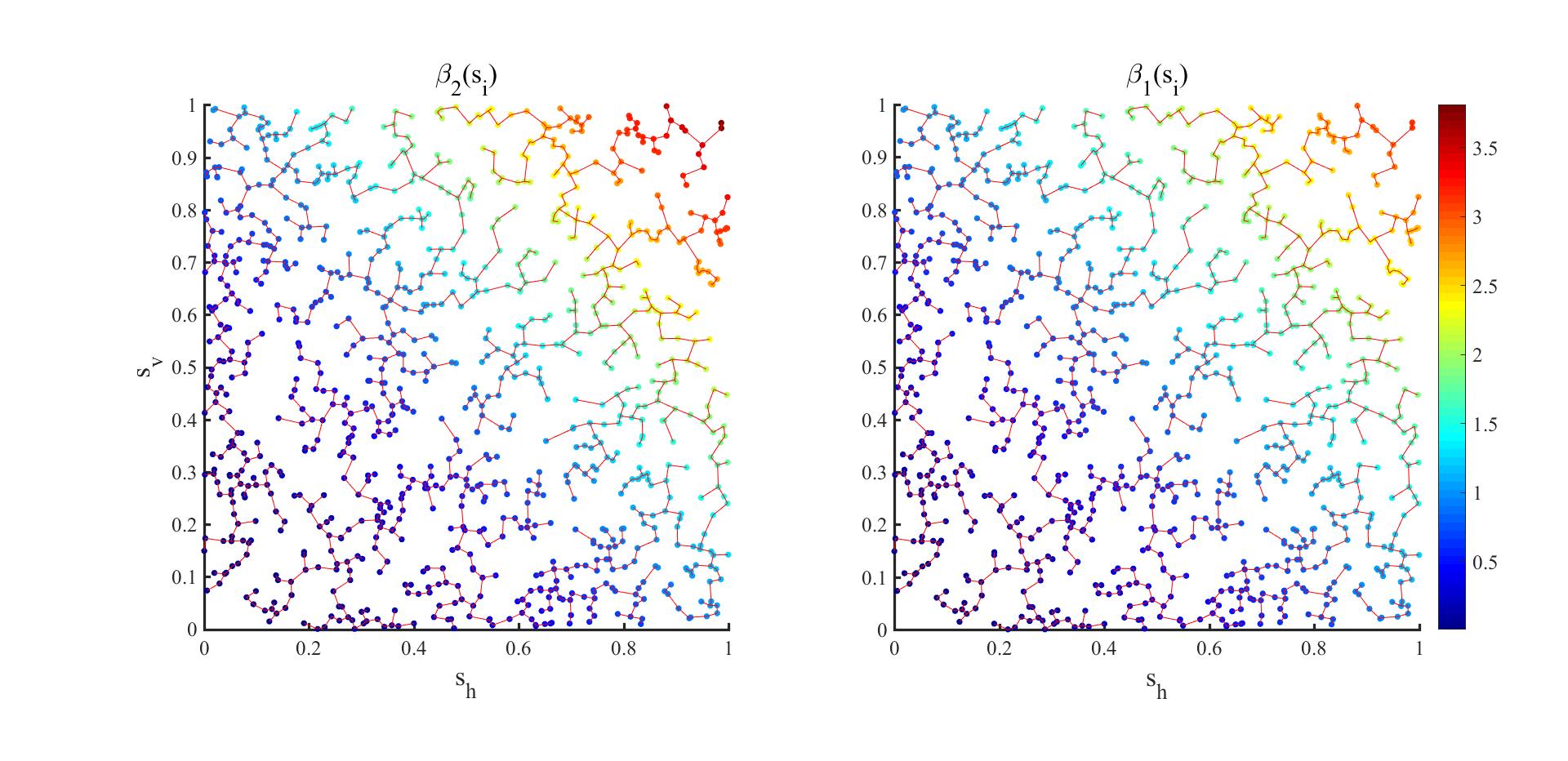}
	\centering
	\caption{{The points/colors represent the locations/coefficient values, and the solid lines represent the edges in MST. For $\bm{s}=(s_h, s_v)$ and $s_{hv}=s_h+s_v$, $\beta_2(\bm{s})=s_{hv}^2$ and $\beta_1(\bm{s})=s_{hv}^{1.7}$.}}
	\label{ssffigure4}
\end{figure}

{
	The true regression coefficients $\{\beta_k(\bm{s}_i)\}_{k=1}^2$ in this study are smooth over the whole region, i.e.,  the assumption made in GWR and PSE holds, see  Figure \ref{ssffigure4}.  Other settings are the same as those in Section 4 of the main paper.

	Table \ref{sstable3} summarizes the results of the five methods.  SCVC performs slightly worse than PSE and better than other methods. This is because the assumption made in PSE holds under this setting. It is worth to point out that PSE performs much better than GWR, although the assumption in GWR also holds under this setting. One possible explanation is that the PSE is based on a global smoothing method, which utilizes all the information of samples, however the GWR is based on a local smoothing method, and only a small number of samples are used for estimation. 
}

\begin{table}[htbp] 
	\small
	\renewcommand\arraystretch{0.8}
	\setlength{\abovecaptionskip}{0pt}
	\setlength{\belowcaptionskip}{10pt}
	\begin{threeparttable}[b]
		\caption{Summary of results for smooth-varying coefficients.} 
		\begin{tabular}{p{2cm}<{\centering}p{1.5cm}<{\centering}p{1.5cm}<{\centering}p{1.5cm}<{\centering}p{1.3cm}<{\centering}p{1.3cm}<{\centering}p{1cm}<{\centering}p{1cm}<{\centering}} 
			\hline
			\hline
			\multicolumn{1}{p{2cm}<{\centering}}{Correlation} & \multicolumn{1}{p{1.5cm}<{\centering}}{Methods}& \multicolumn{1}{p{1.5cm}<{\centering}}{$\text{MSE}_{\beta_2}$}&\multicolumn{1}{p{1.5cm}<{\centering}}{$\text{MSE}_{\beta_1}$}&\multicolumn{1}{p{1.25cm}<{\centering}}{$\text{RI}_2$}&\multicolumn{1}{p{1.25cm}<{\centering}}{$\text{RI}_1$}&\multicolumn{1}{p{1.25cm}<{\centering}}{$\text{IC}_2$} &\multicolumn{1}{p{1.25cm}<{\centering}}{$\text{IC}_1$}\\
			\cmidrule(lr){3-4}\cmidrule(lr){5-6}\cmidrule(lr){7-8}
			&SCVC&0.004&0.006&100&100&1.00&1.00\\
			&&(0.000)&(0.000)&(0.00)&(0.00)&(0.00)&(0.00)\\
			&SCC&0.204&0.231&1.47&1.94&176.98&126.07\\
			&&(0.002)&(0.003)&(0.02)&(0.03)&(1.37)&(1.31)\\
			weak&SCC*&0.231&0.259&3.15&5.44&57.88&34.31\\
			&&(0.001)&(0.001)&(0.01)&(0.04)&(0.15)&(0.11)\\
			&GWR&0.112&0.151&-&-&-&-\\
			&&(0.001)&(0.002)&-&-&-&-\\
			&PSE&0.004&0.005&-&-&-&-\\
			&&(0.000)&(0.000)&-&-&-&-\\
			\hline
			&SCVC&0.022&0.069&94.28&100&1.20&1.00\\
			&&(0.001)&(0.002)&(1.57)&(0.00)&(0.05)&(0.00)\\
			&SCC&1.390&2.513&1.97&11.49&148.32&48.05\\
			&&(0.015)&(0.025)&(0.04)&(0.29)&(1.78)&(0.88)\\
			strong&SCC*&1.246&2.469&2.61&14.22&63.45&27.43\\
			&&(0.005)&(0.010)&(0.01)&(0.07)&(0.19)&(0.10)\\
			&GWR&0.940&1.783&-&-&-&-\\
			&&(0.007)&(0.013)&-&-&-&-\\
			&PSE&0.019&0.052&-&-&-&-\\
			&&(0.001)&(0.002)&-&-&-&-\\
			\hline
			\hline
		\end{tabular}
		\label{sstable3}
		\begin{tablenotes} 
			\item SCVC: spatially clustered varying coefficient method; SCC: spatially clustered coefficient regression based on LASSO; SCC*: spatially clustered coefficient regression based on SCAD; GWR: geographically weighted regression; PSE: $P$-spline estimator. $\text{MSE}_{\beta_k}$/$\text{RI}_k$/$\text{IC}_k$: mean squared error $(\times 10)$/rand index $(\times 100)$/number of identified clusters, for $k$-th covariate, $k=1, 2$. Values in the parentheses are the standard errors. Note that GWR and PSE can not identify clusters.
		\end{tablenotes}
	\end{threeparttable}
\end{table}

\vspace{-0.9cm}
\small
\setlength{\bibsep}{0.4ex}
\bibliographystyle{Chicago}
\bibliography{Bibliography-MM-MC}
\end{document}